\documentclass[12pt,epsf]{article}
\usepackage{verbatim}
\usepackage{anyfontsize}
\usepackage{dsfont}
\usepackage{tikz}
\usepackage{sidecap}
\sidecaptionvpos{figure}{t}
\usepackage{subfigure}
\usepackage{enumitem}
\usepackage[english]{babel}
\usepackage{enumitem}  

\usepackage{amssymb,amsmath}
\usepackage{graphicx, xcolor, varwidth}
\usepackage{setspace}
\usepackage[permil]{overpic}
\usepackage{cite}
\usepackage{mathtools}

\DeclareSymbolFont{matha}{OML}{txmi}{m}{it}
\DeclareMathSymbol{v}{\mathord}{matha}{118}

\usepackage[framemethod=default]{mdframed}
\newmdenv[skipabove=7pt,
skipbelow=7pt,
rightline=false,
leftline=false,
topline=false,
bottomline=false,
backgroundcolor=blue!10,
linecolor=blue,
innerleftmargin=5pt,
innerrightmargin=5pt,
innertopmargin=5pt,
innerbottommargin=5pt,
leftmargin=0cm,
rightmargin=0cm,
linewidth=4pt]{bBox}

\colorlet{darkblue}{blue!70!black}
\colorlet{darkgreen}{green!70!black}

\usepackage[colorlinks=true,urlcolor=darkblue,linktocpage=true,linkcolor=darkblue,citecolor=darkblue]{hyperref}

\numberwithin{equation}{section}
\DeclareMathSymbol{v}{\mathord}{matha}{118}
\newcommand{\be}{\begin{equation}}
\newcommand{\ee}{\end{equation}}
\newcommand{\bea}{\begin{eqnarray}}
\newcommand{\eea}{\end{eqnarray}}
\newcommand{\bear}{\begin{eqnarray}}
\newcommand{\eear}{\end{eqnarray}}
\newcommand{\beas}{\begin{eqnarray*}}

\newcommand{\eeas}{\end{eqnarray*}}
\newcommand{\ba}{\begin{array}}
\newcommand{\ea}{\end{array}}

\def\ba#1\ea{\begin{align}#1\end{align}}
\def\bs#1\es{\begin{split}#1\end{split}}

\renewcommand{\r}{\rho}
\newcommand{\br}{\bar{\rho}}

\newcommand{\pd}[2][1]{\ifnum#1=1 \frac{\partial}{\partial {#2}} \else
  \frac{\partial^#1}{\partial {#2}^{#1}}\fi}
\newcommand{\dpd}[2][1]{\ifnum#1=1 \dfrac{\partial}{\partial {#2}} \else
  \frac{\partial^#1}{\partial {#2}^{#1}}\fi}
\newcommand{\td}[2][1]{\ifnum#1=1 \frac{d}{d{#2}} \else
  \frac{d^#1}{d{#2}^{#1}}\fi}

\renewcommand{\(}{\left(}
\renewcommand{\)}{\right)}

\newcommand{\nbox}{{\,\lower0.9pt\vbox{\hrule \hbox{\vrule height 0.2 cm \hskip 0.19 cm \vrule height 0.2 cm}\hrule}\,}}

\def\O{{\cal O}}

\newcommand{\bz}{\bar{z}}

\newcommand{\CFTUV}{\text{CFT}_{\text{UV}}}
\newcommand{\CFTIR}{\text{CFT}_{\text{IR}}}



\newcommand{\z}{\mathfrak{z}}

\begin{document}
\begin{spacing}{1.3}
\begin{titlepage}

\begin{center}
{\Large \bf 

Renormalization Group Flows, the $a$-Theorem and  Conformal Bootstrap

}

\vspace*{6mm}

Sandipan Kundu

\vspace*{6mm}

\textit{Department of Physics and Astronomy,
\\ Johns Hopkins University,
Baltimore, Maryland, USA\\}

\vspace{6mm}

{\tt \small kundu@jhu.edu}

\vspace*{6mm}
\end{center}

\begin{abstract}
Every renormalization group flow in $d$ spacetime dimensions can be equivalently described as spectral deformations of a generalized free CFT in $(d-1)$ spacetime dimensions. This can be achieved by studying the effective action of the Nambu-Goldstone boson of broken conformal symmetry in anti-de Sitter space and then taking the flat space limit. This approach is particularly useful in even spacetime dimension where the change in the Euler anomaly $ a_{UV}-a_{IR}$ can be related to anomalous dimensions of lowest twist multi-trace operators in the dual CFT. As an application, we provide a simple proof of the 4d $a$-theorem using the dual description. Furthermore, we reinterpret the statement of the $a$-theorem in 6d as a conformal bootstrap problem in 5d.

\end{abstract}

\end{titlepage}
\end{spacing}

\vskip 1cm
\setcounter{tocdepth}{2}  
\tableofcontents

\begin{spacing}{1.3}

\section{Introduction}

Every renormalization group (RG) flow can be described as spontaneous breaking of conformal symmetry of some conformal field theory (CFT). This provides an elegant formalism to study general features of RG flows in terms of the effective action of a massless {\it dilaton}, which is the Nambu-Goldstone boson of  spontaneously broken conformal symmetry \cite{Komargodski:2011vj}. In this paper, we view the flat space dilaton effective theory as a theory in  anti-de Sitter (AdS) space with finite but large radius $R_{\text{AdS}}$ and then take the flat space limit $R_{\text{AdS}}\rightarrow \infty$. We know from the conformal bootstrap that scalar effective field theories inside an AdS ``box" have the advantage of a dual CFT description.  To be specific, it has been established by \cite{Heemskerk:2009pn} and subsequent authors  \cite{Heemskerk:2010ty,Fitzpatrick:2010zm,Penedones:2010ue,ElShowk:2011ag,Fitzpatrick:2011ia,Fitzpatrick:2011hu,Fitzpatrick:2011dm,Fitzpatrick:2012cg,Goncalves:2014rfa,Alday:2014tsa,Hijano:2015zsa,Aharony:2016dwx}, that scalar effective field theories in AdS in $d$ dimensions are in one-to-one correspondence with perturbative solutions of crossing symmetry in CFT  in $(d-1)$ dimensions. This connection enables us to analyze the dilaton effective theory in $d\ge3$ dimensions using methods from the conformal bootstrap in $(d-1)$ dimensions. 

For example, a free scalar theory in AdS enjoys a dual description in terms of a {\it generalized free CFT} of a scalar primary $\O$ of dimension $\Delta_\O$. Of course, this dual CFT is required by crossing symmetry to contain infinite towers of $N$-trace operators with spin $\ell$ and dimensions $N\Delta_\O+2n+\ell$, for all non-negative integer $n$, which we denote as $[\O^N]_{n,\ell}$. Besides, each interaction of the scalar field in AdS corresponds to a specific perturbative solution to crossing symmetry in the dual CFT. Thus, the dilaton effective theory in AdS$_d$ can be equivalently described as a CFT$_{d-1}$ which is obtained by deforming operator dimensions and OPE coefficients of  a generalized free theory.

\begin{figure}
\centering
\includegraphics[scale=0.33]{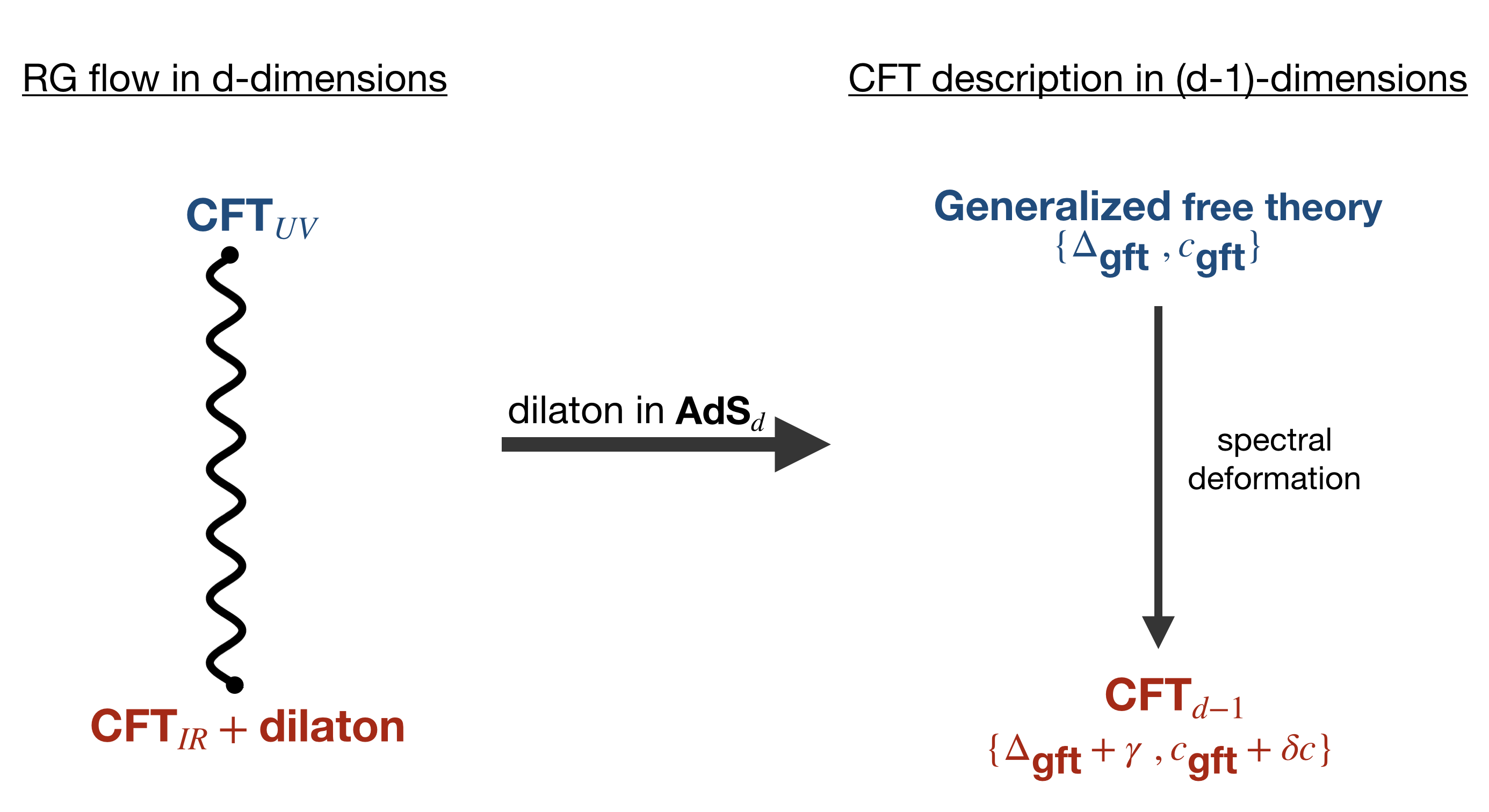}
\caption{ \label{intro} \small Every RG flow connecting two conformal fixed points $\CFTUV$ and $\CFTIR$ in $d$ spacetime dimensions can be equivalently described as a CFT in $(d-1)$ spacetime dimensions for $d\ge 3$. The CFT$_{d-1}$ is obtained by deforming operator dimensions and OPE coefficients of  a $(d-1)$ dimensional generalized free theory.}
\end{figure}

The above discussion implies that every RG flow connecting two conformal fixed points in $d$ dimensions can be interpreted as  deformations of the spectrum of a generalized free CFT$_{d-1}$ for $d\ge 3$, as shown in figure \ref{intro}. This dual CFT$_{d-1}$, for any unitary RG flow, must obey the Euclidean axioms. Hence, general aspects of unitary RG flows, such as {\it irreversibility}, can be studied completely within the paradigm of conformal bootstrap in one lower dimension. This philosophy parallels recent developments in S-matrix bootstrap where conformal bootstrap methods were used to study quantum field theory (QFT) in AdS \cite{Paulos:2016fap,Paulos:2016but,Paulos:2017fhb,Homrich:2019cbt}.

The irreversibility of RG flows is of fundamental importance in QFT. Consider a unitary RG flow where $\CFTUV$ flows to $\CFTIR$. Since RG flows represent coarse-graining, it is expected that  fundamental principles such as quantum mechanics and symmetries should forbid any RG flow that starts from $\CFTIR$ and ends at $\CFTUV$. The first concrete realization of this expectation was the  C-theorem due to Zamolodchikov (1986) which established the irreversibility of RG flows for 2d QFTs \cite{Zamolodchikov:1986gt}. In 1988, Cardy proposed a natural generalization of the 2d theorem to any even spacetime dimensions in terms of the Euler central charge $a$ \cite{Cardy:1988cwa}. In $d$ (even) spacetime dimensions, the conformal trace anomaly has the following structure \cite{Deser:1976yx,Duff:1977ay,Fradkin:1983tg,Deser:1993yx}
 \be\label{anomaly0}
 \langle T^\mu_\mu\rangle =-(-1)^{d/2} a\ E_d+ \sum_{i} c^{(i)} I_i\ 
 \ee
up to total derivative terms which can be removed by adding finite and covariant counter-terms in the effective action. Here, $E_d$ is the $d$-dimensional Euler density and $I_i$ are local Weyl invariants of conformal weight $d$. Cardy conjectured that the Euler central charge $a$ decreases under unitary RG flows
\be
a_{UV}\ge a_{IR}
\ee
implying the irreversibility in even spacetime dimensions \cite{Cardy:1988cwa}. In 4d, there was ample evidence in favor of this conjecture, however, general proof of the  $a$-theorem was an open problem for over twenty years until an elegant proof was found by Komargodski and Schwimmer in 2011  \cite{Komargodski:2011vj} (see also \cite{Komargodski:2011xv}). On the other hand, the 6d $a$-theorem has resisted all attempts at proof. This is particularly surprising since the dilaton based formalism of \cite{Komargodski:2011vj,Komargodski:2011xv} does extend to 6d \cite{Maxfield:2012aw,Elvang:2012st}. Moreover, there is strong evidence for the 6d $a$-theorem for flows that preserve supersymmetry \cite{Cordova:2015fha,Cordova:2015vwa,Heckman:2015axa,Heckman:2016ssk,Heckman:2018jxk}. However,  as it was discussed in \cite{Maxfield:2012aw,Elvang:2012st,Grinstein:2014xba, Grinstein:2015ina}, there are several major obstructions to a general proof of the $a$-theorem in 6d. The purpose of this paper is to interpret the 6d $a$-theorem as a CFT$_5$ problem which may yield to conformal bootstrap techniques.

The proof of the 4d $a$-theorem in \cite{Komargodski:2011vj,Komargodski:2011xv}  follows from the unitarity and analyticity of the dilaton four-point amplitude. The dual CFT$_3$ description of a 4d RG flow also provides a simple proof of the $a$-theorem. In this approach, the spin-2 lowest twist operator $[\O^2]_{0,2}$, where $\O$ is a scalar primary dual to the dilaton, acquires an anomalous dimension $\gamma_2$ under the RG flow implying 
\be
a_{UV}- a_{IR}=-\tilde{\Delta}_f^4\ \gamma_2\ .
\ee
In the above relation, the gap $\tilde{\Delta}_f$ is a CFT cut-off scale which is determined by the details of conformal symmetry breaking that triggers the RG flow. The $a$-theorem now simply follows from the CFT Nachtmann theorem \cite{Komargodski:2012ek,Costa:2017twz,kundu}, as well as from causality \cite{Hartman:2015lfa} which requires  $\gamma_2 \le 0$. We also construct a monotonically decreasing CFT$_3$ function that interpolates between $a_{UV}$ and $a_{IR}$.

Similarly, the 6d $a$-theorem can be rephrased as a CFT$_5$ statement\footnote{Note that in 6d a CFT in the UV can flow to a fixed point which is scale-invariant but non-conformal \cite{Cordova:2015fha}. In this paper, we will only consider RG flows between two conformal fixed points. } 
\be\label{intro1}
a_{UV}- a_{IR}={\tilde{\Delta}_f}^{12}\(-\delta \gamma_3+\alpha \gamma_2^2\)\ , 
\ee
where, $\gamma_2$ and $\gamma_3$ are the anomalous dimensions of the lowest twist spin-2 and spin-3 operators respectively and $\alpha$ is a real, model-independent universal numerical factor which is completely fixed by symmetry (see section \ref{section_atheorem} and \ref{section_conclusions}). Clearly, in the dual CFT$_5$ the double-trace operator $[\O^2]_{0,2}$ is the lowest twist spin-2 operator. Whereas,  the lowest twist spin-3 operator is the triple-trace operator $[\O^3]_{0,3}$. Anomalous dimensions of triple-trace operators are complicated objects simply because three-particle bound states are complicated. As a result $\gamma_3$ receives contributions from various different processes. So, we have defined a subtracted anomalous dimension $\delta \gamma_3$ which corresponds to the binding energy of a spin-3 three-particle state in AdS$_6$ arising from purely three-particle interactions.\footnote{A part of $\gamma_3$ comes entirely from the anomalous dimension of the double-trace operator $[\O^2]_{0,2}$. We define $\delta \gamma_3$ by subtracting this contribution. In particular, at the leading order the exact relation is $\delta \gamma_3=\gamma_3-\frac{51}{22}\gamma_2$. For a detail discussion see section \ref{section_atheorem}.}

To be specific, the relation (\ref{intro1}) holds for 6d RG flows from spontaneously broken conformal symmetry. However, a more standard scenario in which  RG flows  are triggered by adding  relevant (or marginally relevant) operators  that break  conformal symmetry explicitly can be thought of as a special case with $\gamma_2^2\ll |\delta \gamma_3|$. Hence, the above relation simplifies further for explicitly broken conformal symmetry 
\be\label{intro2}
a_{UV}- a_{IR}=-{\tilde{\Delta}_f}^{12} \delta\gamma_3\ .
\ee

Anomalous dimensions of odd spin operators do obey a generalized Nachtmann theorem \cite{kundu} that provides a lower bound on $\gamma_3$, however, we are not aware of any CFT theorem that implies $\delta\gamma_3 \le \alpha \gamma_2^2$.  Thus,  the 6d $a$-theorem never ceases to be a difficult problem. Nonetheless, the relations (\ref{intro1}) and (\ref{intro2}) suggest that the $a$-theorem now can be explored using numerical bootstrap techniques. This is indeed encouraging since spectral deformations of generalized free theories are sufficiently simple  to be analyzed numerically. The hope is that an upper bound on $\gamma_3$ can be obtained from the numerical bootstrap which will settle the 6d $a$-theorem once and for all.

The outline of this paper is as follows.  In section \ref{section_RG} we review the dilaton effective theory associated with RG flows in 6d. In section \ref{section_CFT} we study this dilaton effective action in AdS and discuss its dual description in terms of spectral deformations of a generalized free CFT$_5$. We then utilize the CFT$_5$ description to relate the change in the Euler anomaly $\Delta a$ under the RG flow to anomalous dimensions of lowest twist multi-trace CFT operators in section \ref{section_atheorem}. Finally we summarize our conclusions in section \ref{section_conclusions}.

\section{RG flows in six dimensions}\label{section_RG}
\addtocontents{toc}{\protect\setcounter{tocdepth}{-1}}
In this paper we will mainly focus on RG flows between conformal fixed points in 6d. We will also comment on some aspects of  4d RG flows throughout the paper.  In fact, RG flows in 4d can be thought of a simpler version of the 6d case. 

The trace of the stress tensor for 6d CFTs is anomalous in the presence of a background metric. The trace anomaly can be characterized by 4 central charges $\{a, c^{(1)},c^{(2)},c^{(3)}\}$ where $a$ is the Euler central charge and  $c^{(i)}$ are central charges associated with 3 Weyl invariants (see appendix \ref{app_anomaly} for details). Central charges $c^{(i)}$ also appear in the stress tensor three-point function and hence they are constrained by the conformal collider bounds \cite{Hofman:2008ar}. On the other hand, there are no constraints on the Euler central charge $a$ since it only  contributes to the stress tensor four-point function. In contrast, the Euler anomaly does obey a positivity condition in 2d and 4d. Hence, the claim that the Euler central charge is a measure of the effective number of degrees of freedom is slightly stronger than the $a$-theorem in 6d.

\subsection{Spontaneously broken conformal symmetry}
Consider a $\CFTUV$ in (5+1)-dimensions with the Euler central charge $a_{UV}$. We assume that the $\CFTUV$ has a moduli space of vacua. This enables us to break the conformal symmetry spontaneously by turning on VEVs for an operator $O$. The VEV $\langle O\rangle\sim f$ emanates an RG flow that leads to some $\CFTIR$. In addition, the Nambu-Goldstone theorem requires that the spontaneously broken conformal symmetry generates a {\it massless} boson -- the dilaton.  So, in general the low energy theory consists of $\CFTIR$ and a massless dilaton $\tau$
 \be\label{IR}
 S_{\text{IR}}=\CFTIR+ S_{\text eff}[\tau]\ ,
 \ee
 however, in certain cases the $\CFTIR$ can be trivial. The dilaton effective action $S_{\text eff}[\tau]$ is highly constrained even in 6d. This becomes obvious when we couple the theory to a metric $g_{\mu\nu}(x)$ and study the variation of the action under diff$\times$Weyl transformations, where Weyl transformations are defined as
 \be\label{weyl}
 g_{\mu\nu}(x)\rightarrow e^{2\sigma(x)}g_{\mu\nu}(x)\ , \qquad \tau(x)\rightarrow \tau(x)+\sigma(x)\ .
 \ee

\begin{figure}
\centering
\includegraphics[scale=0.30]{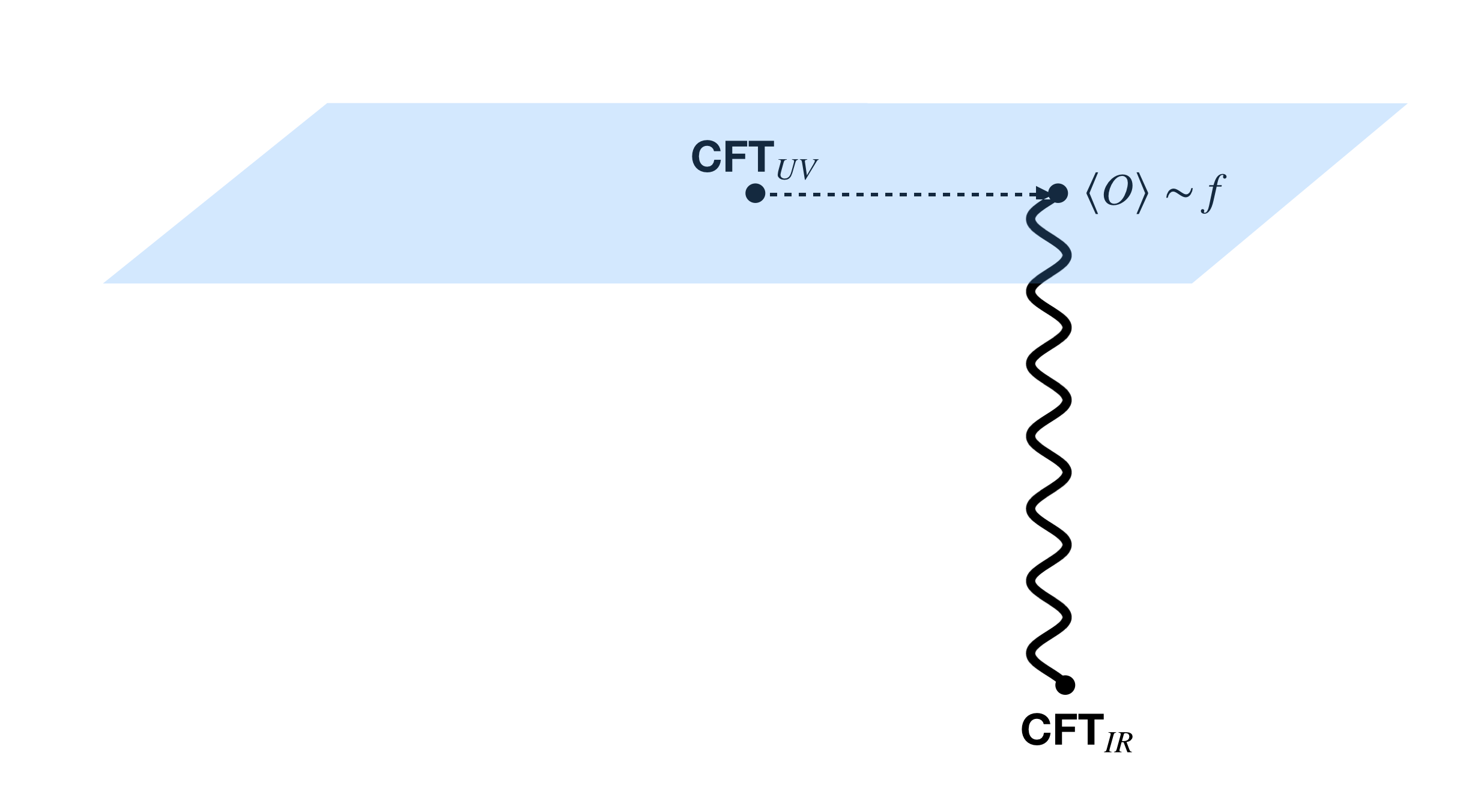}
\caption{ \label{rg} \small Conformal symmetry of $\CFTUV$ is spontaneously broken by turning on a VEV $\langle O\rangle\sim f$. As a result the theory flows to a low energy theory consists of a $\CFTIR$ and a massless dilaton.}
\end{figure}

The IR theory (\ref{IR}) must have the same anomalies as the UV theory $\CFTUV$. This follows from the fact that in flat space the stress tensor remains traceless as an operator $T^\mu_\mu=0$ even when the conformal symmetry is spontaneously broken. The  standard anomaly matching arguments of \cite{Schwimmer:2010za} then imply that the total IR anomalies should match the total UV anomalies. This requirement, as shown in \cite{Komargodski:2011vj, Maxfield:2012aw,Elvang:2012st}, completely fixes the low energy dilaton effective action $S_{\text eff}[g_{\mu\nu},\tau]$.\footnote{For a $d$-dimensional generalization see \cite{Elvang:2012yc}.} The flat space limit of $S_{\text eff}[g_{\mu\nu}=\eta_{\mu\nu},\tau]$ then leads to a simple yet non-trivial effective action $S_{\text eff}[\tau]$. 

Of course, in general $\CFTUV$ and $\CFTIR$ have different anomalies. Hence, all changes in anomalies  in the flow from  $\CFTUV$ to $\CFTIR$ must be compensated by the dilaton. This completely fixes the Weyl variation of the dilaton effective action
\be\label{variation}
\delta_\sigma S_{\text eff}[g_{\mu\nu},\tau]=\int d^6x \sqrt{-g} \sigma(x)\left(\Delta a E_6+ \sum_{i=1}^3 \Delta c^{(i)} I_i \right)
\ee
with $\Delta a\equiv a_{UV}-a_{IR}$ and $\Delta c^{(i)}\equiv c^{(i)}_{UV}-c^{(i)}_{IR}$. Anomalies $\{a_{IR},c^{(i)}_{IR}\}$ should be understood as the total anomalies of $\CFTIR$ and the massless dilaton. The variational equation (\ref{variation}) can now be solved systematically to obtain $S_{\text eff}[g_{\mu\nu},\tau]$ which we review next following \cite{Elvang:2012st}. 

One obvious way the equation (\ref{variation}) can be simplified by writing  
\be\label{simple}
S_{\text eff}[g_{\mu\nu},\tau]=\int d^6x \sqrt{-g} \tau(x)\left(\Delta a E_6+ \sum_{i=1}^3 \Delta c^{(i)} I_i \right)+S_{nl}+S_{inv}\ .
\ee
Note that $I_i$'s are invariant under Weyl transformations, however, the Euler density is not. Hence, the first term in the above equation  generates the correct Weyl variation (\ref{variation}) plus an extra term $\Delta a\int d^6x \sqrt{-g} \tau(x)\delta_\sigma E_6$. This extra contribution is cancelled by adding  a non-linear action $S_{nl}$ of $\tau$. In addition, we can add any diff$\times$Weyl invariant action $S_{inv}$ without affecting (\ref{variation}). The main advantage of writing $S_{\text eff}[g_{\mu\nu},\tau]$ as (\ref{simple}) is that the non-linear action $S_{nl}$ is completely fixed by the UV and the IR fixed points of the RG flow modulo diff$\times$Weyl invariant terms. Furthermore, the linearity of (\ref{variation}) implies that $S_{nl}$ is unique as well up to invariant terms. These properties, as shown in  \cite{Elvang:2012st},  are enough to determine $S_{nl}$ exactly. In particular, in the flat space limit, $S_{nl}$ is given by a simple formula \cite{Elvang:2012st}
\be\label{uni}
S_{nl}|_{g_{\mu\nu}=\eta_{\mu\nu}}=3\Delta a\int d^6x \tau \Box^3 \tau + \cdots\ ,
\ee
where, dots represent terms that can be absorbed in $S_{inv}$. The {\it universality} of $S_{nl}$ makes this simple dilaton-based approach a rather powerful tool to study general features of RG flows in even dimensions.

Let us now focus on $S_{inv}$. This is the part of the dilaton effective action which is {\it non-universal}. In physical systems any kind of universality is of significance only when non-universal effects are also highly constrained because of some symmetries. This is precisely the case with $S_{inv}$. The diff$\times$Weyl invariance implies that at each derivative order only a finite number of terms can appear in $S_{inv}$. Moreover, only a few of these terms are expected to survive after we take the flat space limit. Both of these conditions can be efficiently implemented by constructing $S_{inv}$ only from non-vanishing curvature invariants of the Weyl-invariant metric
\be\label{metric0}
\hat{g}_{\mu\nu}=e^{-2\tau}\eta_{\mu\nu}\ .
\ee
The fact that $S_{inv}$ is non-universal is encoded in coefficients of these curvature invariants which depend on details of the RG flow.  Since 
$S_{nl}$ has six derivatives, we need to consider all Weyl-invariants with maximum of six derivatives to construct $S_{inv}$.  Up to six derivatives, there are only six independent non-vanishing Weyl-invariants  and the most general $S_{inv}$ is given by \cite{Elvang:2012st}
\begin{align}\label{inv}
S_{inv}|_{g_{\mu\nu}=\eta_{\mu\nu}}=\int d^6x \sqrt{-\hat{g}}\left(-\frac{f^4}{10} \hat{R}-\frac{\hat{b} f^2}{2}\hat{R}^{\mu\nu}\hat{R}_{\mu\nu}+ b' f^2 \hat{R}^2+b_1\hat{R}^3+b_2 \hat{R}\hat{R}^{\mu\nu}\hat{R}_{\mu\nu}+b_3 \hat{R}\hat{\Box}\hat{R} \right)
\end{align}
where, $\hat{R}$ and $\hat{R}^{\mu\nu}$ are computed using the Weyl-invariant metric (\ref{metric0}). In the above equation $f$ has dimension of mass and $b', \hat{b}, b_i$ are dimensionless coefficients.\footnote{Note that our $\hat{b}=b/f^2$ of \cite{Elvang:2012st}. }  Numerical factors are chosen for later convenience. This Weyl-invariant action can be further simplified by using the equation of motion of $\tau$. The last four terms of (\ref{inv}) vanish once we impose the on-shell condition for the dilaton and hence these terms can only affect low energy observables at subleading orders.\footnote{As in \cite{Komargodski:2011vj, Elvang:2012st} there is no cosmological constant term in (\ref{inv}) for spontaneously broken conformal symmetry. For explicit breaking, the flow in general can generate a cosmological constant term in the IR, however, we will always tune the IR cosmological constant term to zero by adding a suitable counterterm.}

Finally, we are ready to write down the low-energy effective action for the dilaton by taking the flat space limit of (\ref{simple}). In the flat space limit $E_6$ and $I_i$'s vanish and hence the dilaton effective action only knows about  $a_{UV}-a_{IR}$ through $S_{nl}$. Putting everything together, $S_{\text{eff}}[\tau]$ is given by \cite{Elvang:2012st}
\be\label{dilaton}
S_{\text{eff}}[\tau]=\int d^6x\left(-2f^4 (\partial \tau)^2 e^{-4\tau}+4\hat{b}f^2 e^{-\tau}\Box^2 e^{-\tau} +3\Delta a\tau \Box^3 \tau\right) \ .
\ee
The above effective action is deceptively simple. Just like the 4d case, the non-canonical kinetic term is completely fixed by the constant $f$ which is related to the VEV of the  operator $O$ of $\CFTUV$ which triggers the RG flow. Unfortunately the similarity ends here. Unlike the 4d case, the coefficient of the 4-derivative term in (\ref{dilaton}) depends on the details of the RG flow. On the other hand, the 6-derivative term in (\ref{dilaton}) is universal which suggests that the dilaton-based approach can eventually lead to a proof of the a-theorem in 6d. 

There are two major obstructions to a proof of the 6d a-theorem. First of all, the non-universality of the 4-derivative term implies that the universality  of the 6-derivative term is of limited use. Indeed, it is rather difficult to find an observable that receives dominant contribution only from the 6-derivative term of the action (\ref{dilaton}). Secondly, any such observable, if found, can only lead to a proof if that observable satisfies some strict positivity condition which follows from general principles such as unitarity or causality. In general, energy conditions are rather rare in QFT which explains why the a-theorem in 6d is so elusive.

\subsection{Physical dilaton}

The effective action (\ref{dilaton}) has a simple form. However, the dilaton field $\tau$ is not very useful when we study the theory using traditional tools of QFT. This issue can be easily resolved by a simple field redefinition: 
\be
e^{-2\tau}=1-\frac{\tilde{\phi}}{f^2}\ , \qquad \tilde{\phi}=\phi-\frac{\hat{b}}{f^2}\Box \phi+\frac{6\hat{b}^2-3\Delta a}{4f^4}\Box^2 \phi+\O\left(\frac{\Box^3}{f^6} \right)\ ,
\ee
where, the {\it physical} dilaton field $\phi$ has a canonical kinetic term. Of course, this comes at a price. The resulting action 
\be\label{final}
S_{\text{eff}}[\phi]=\int d^6x \left(-\frac{1}{2}(\partial \phi)^2+ \mathcal{L}_{\phi^3}+\mathcal{L}_{\phi^4}+\mathcal{L}_{\phi^5}+\mathcal{L}_{\phi^6}+\cdots \right)\ ,
\ee
where $ \mathcal{L}_{\phi^n}$ represents $n-\phi$ interaction, does not possess the apparent simplicity of (\ref{dilaton}). The three-$\phi$ interaction is trivial 
\be\label{int3}
\mathcal{L}_{\phi^3}=\frac{\hat{b}}{2f^4}\phi^2 \Box^2 \phi+\frac{3\Delta a}{4f^6}\phi^2 \Box^3 \phi\ 
\ee
and it only contributes to exchange (or loop) diagrams. More interesting higher-point interactions are given by
\begin{align}
\mathcal{L}_{\phi^4}=&\frac{\hat{b}}{f^6}\left( \frac{1}{4}\phi^3 \Box^2 \phi +\frac{1}{16} \phi^2 \Box^2 \phi^2\right)+\frac{\Delta a}{f^8}\left( \frac{1}{2}\phi^3 \Box^3 \phi +\frac{3}{16} \phi^2 \Box^3 \phi^2\right)\ ,\label{int4}\\
\mathcal{L}_{\phi^5}=&\frac{\hat{b}}{32f^8}\left(5 \phi^4 \Box^2 \phi +2 \phi^3 \Box^2 \phi^2\right)+\frac{\Delta a}{8f^{10}}\left( 3\phi^4 \Box^3 \phi +2 \phi^3 \Box^3 \phi^2\right)\ ,\\
\mathcal{L}_{\phi^6}=&\frac{\hat{b}}{128f^{10}}\left(14 \phi^5 \Box^2 \phi+5 \phi^4 \Box^2 \phi^2+2 \phi^3 \Box^2 \phi^3\right)\nonumber\\
&\qquad \qquad +\frac{\Delta a}{240f^{12}}\left( 72 \phi^5 \Box^3 \phi+45 \phi^4 \Box^3 \phi^2+20 \phi^3 \Box^3 \phi^3\right)\ \label{int6}
\end{align}
up to total derivative terms that do not contribute to correlators. To summarize, we started with a well behaved UV theory $\CFTUV$ which flows to a low energy theory consists of a $\CFTIR$ and a massless dilaton with the action (\ref{final}). The full IR theory must be Lorentz invariant, unitary, and causal. This imposes further restrictions on the dilaton effective action (\ref{final}).

Similar to the proof of the 4d a-theorem, we first study the 4-point on-shell scattering amplitude $\mathcal{A}(s,t)$ of the dilaton.\footnote{$s$ and $t$ are the usual the Mandelstam variables.} At low energies, the amplitude is dominated by the $\phi^2 \Box^2 \phi^2$ term of the dilaton effective action. The analyticity property of the 4-point scattering amplitude $\mathcal{A}(s,t)$ implies that the parameter $\hat{b}$ obeys a sum rule \cite{Adams:2006sv}
\be\label{sumrule}
\hat{b}=\frac{2f^6}{\pi}\int_0^\infty ds \frac{\mbox{Im}\ \mathcal{A}(s,0)}{s^3} >0\ ,
\ee
where the positivity condition follows from unitarity: $\mathcal{A}(s,0)=s \sigma_{tot}(s)>0$.  The derivation of this dispersive sum-rule, as correctly pointed out in \cite{Elvang:2012st}, requires an additional assumption that the 4-point on-shell scattering amplitude $\mathcal{A}(s,t)$ grows slower than $s^2$ for large $s$. In 6d, there is no rigorous QFT argument that validates this assumption, however, there are strong reasons to believe that this positivity condition is still true. This expectation is also supported by the classical causality based argument of \cite{Adams:2006sv}.

On the other hand, we still do not have a general proof of the $a$-theorem in 6d in spite of the heroic attempt by the authors of \cite{Elvang:2012st}. The dilaton based approach apparently is not as powerful in 6d as in 4d. This is clearly visible even at the level of the effective action (\ref{final}). Consider any $n$-point on-shell scattering amplitude of the dilaton. At low energies, any such amplitude is clearly dominated by $\hat{b}$. Hence, any dispersion relation for $\Delta a$ must involve integral of some non-trivial function of the   $n$-point scattering amplitude  of the dilaton. However, any such dispersion relation has the limitation that its positivity does not immediately follow from unitarity or causality.  This is precisely the reason why 6d $a$-theorem is a hard problem.

\subsection{Explicitly broken conformal symmetry}
We now consider the case where some $\CFTUV$ is deformed by a relevant or marginally relevant operator $M^{6-\Delta}O_{\Delta}$ in 6d. This breaks conformal symmetry explicitly which triggers an RG flow to some $\CFTIR$. At first sight, this scenario appears to be completely different from the scenario where conformal symmetry is broken spontaneously. However,  Komargodski and Schwimmer have argued in \cite{Komargodski:2011vj} that every explicit conformal symmetry breaking can be described  in terms of a spontaneously broken conformal symmetry. The argument is elegant yet simple. Any relevant deformation $M^{6-\Delta}O_{\Delta}$ always introduces an operatorial anomaly to the trace of the stress tensor which spoils the anomaly matching argument. This operatorial anomaly can be conveniently removed by introducing the massless dilaton field as a conformal compensator $\Omega(x)=f^2 e^{-\tau(x)}$ and then replacing $M^2\rightarrow  \frac{M^2}{f^2} \Omega(x)$. In this scenario, $f$ is a free parameter which should be thought of as the decay constant of the dilaton field. The stress tensor of this modified theory is traceless which enables us to describe the explicit symmetry breaking of the original theory as a spontaneous symmetry breaking of the modified theory. In particular, the  explicit symmetry breaking of the $\CFTUV$ now can be implemented by giving the dilaton a VEV: $\langle \Omega\rangle=f^2$. As a result the theory flows to a low energy theory consists of $\CFTIR$ and a dilaton. However, the absence of the operatorial anomaly now guarantees that the total IR anomalies must match the total UV anomalies. Hence, the preceding discussion applies here as well. 

What distinguishes RG flows with explicit symmetry breaking from RG flows with spontaneous symmetry breaking is that the parameter $f$ is completely arbitrary for explicit breaking. So, we can make the interaction between the original theory and the dilaton  weak by choosing $f$ to be much larger than all other mass scales (for example $f\gg M$). In other words, the dilaton can now be treated as a source. For the dilaton effective action (\ref{final}), this effectively means that $| \hat{b}|\ll 1$ for RG flows with explicit symmetry breaking.

\addtocontents{toc}{\protect\setcounter{tocdepth}{1}}
\section{Dilaton effective action and the dual CFT}\label{section_CFT}
\addtocontents{toc}{\protect\setcounter{tocdepth}{-1}}

It is possible that some of the constraints on the flat space dilaton effective action (\ref{final}) from UV consistency are better visible when we place the theory first in AdS$_6$ and then take the flat space limit $R_{\text{AdS}}\rightarrow \infty$. The theory in AdS has a significant advantage. Specifically, it maps the $a$-theorem into a statement about anomalous dimensions in the dual CFT in $(4+1)$-dimensions. Hence, the 6d $a$-theorem can be studied as a conformal bootstrap problem in CFT$_5$, as shown in figure \ref{intro}.

\subsection{Dual CFT}
Consider the  dilaton effective action (\ref{final}) in AdS$_6$ with AdS radius $R_{\text{AdS}}$ large but finite. The action now is simply given by
\be\label{ads}
S_{\text{eff}}[\phi]=\int d^6x \sqrt{g_{AdS}}\left(-\frac{1}{2}g^{\mu\nu}_{AdS}\partial_\mu \phi \partial_\nu \phi+ \mathcal{L}_{int}\right)\ ,
\ee
where, the interactions are obtained from  (\ref{int3}-\ref{int6})
\be\label{int}
\mathcal{L}_{int}=\mathcal{L}_{\phi^3}+\mathcal{L}_{\phi^4}+\mathcal{L}_{\phi^5}+\mathcal{L}_{\phi^6}+\cdots\ .
\ee
For any finite but large $R_{\text{AdS}}$, this theory now can be analyzed by studying its dual CFT$_5$. The dual CFT$_5$, for any unitary RG flow, must be well behaved in the usual sense. In any unitary CFT, analyticity and crossing symmetry of CFT correlators impose non-trivial restrictions on the spectrum. These restrictions in turn constrain interactions of the AdS effective field theory. Of course, these AdS bounds imply analogous bounds for the flat space effective field theory if and only if the CFT description does not breakdown as we take the flat space limit. It is not alway obvious that a smooth flat space limit $R_{\text{AdS}}\rightarrow \infty$ exists for any AdS theory, for example see \cite{Kaplan:2019soo}. However, the fact that all interactions of the dilaton are non renormalizable ensures that a smooth flat space limit of the AdS theory (\ref{ads}) does exist.

The AdS theory (\ref{ads}) does not contain dynamical gravity. This implies that the  stress tensor of the dual CFT$_5$ must decouple from the low energy spectrum. This can be achieved by taking the CFT$_5$ central charge $c_T\rightarrow \infty$, while holding  $f R_{\text{AdS}}\equiv \Delta_{f}$ fixed (but large).\footnote{The central charge $c_T$ is the overall coefficient that appears in the stress tensor two-point function.} The resulting CFT$_5$ should be thought of as an IR effective theory which is well behaved below the cut-off scale $\Delta_f$. This {\it effective} CFT contains a scalar primary operator $\O$ which is dual to the dilaton. Since $\O$ is dual to a Nambu-Goldstone boson, its dimension is completely fixed 
\be
\Delta_\O=5\ 
\ee
implying $\Delta_\O$ does not receive perturbative corrections. 

It is convenient to think of the dual CFT$_5$ as a small perturbation of a generalized free CFT in 5d. When $\mathcal{L}_{int}=0$, the dual CFT$_5$ is exactly a generalized free CFT of the scalar primary $\O$.  In addition, crossing symmetry requires that this generalized free CFT must also contain infinite towers of multi-trace operators $[\O^N]_{n,\ell}$ with spin $\ell$ and dimension $5N+2n+\ell$ for integer $n\ge 0$ \cite{Komargodski:2012ek,Fitzpatrick:2012yx}. Let us now turn on dilaton interactions in AdS. Using conformal bootstrap, it was first shown in \cite{Heemskerk:2009pn} that each interaction in AdS$_d$ corresponds to a perturbative solution of crossing symmetry in the dual CFT$_{d-1}$. In particular, the bulk dilaton theory (\ref{ads}) corresponds to a deformed solution of CFT$_5$ crossing equations in which multi-trace operators $[\O^N]_{n,\ell}$ have dimensions
\be\label{delta}
\Delta^{(N)}(n,\ell)=5N+2n+\ell+\gamma_{n,\ell}^{(N)}\ , \qquad |\gamma_{n,\ell}^{(N)}|\ll 1\ ,
\ee
where, $\gamma_{n,\ell}^{(N)}$ is the anomalous dimension which encodes the information of both $\hat{b}$ and $\Delta a$. We should remark that for large $N$ and $\ell$, there can be multiple distinct multi-trace operators with the same set of quantum numbers $N$, $\ell$ and $n$. For notational convenience, we will denote all of these degenerate operators by the same symbol $[\O^N]_{n,\ell}$.

\subsubsection*{Minimal twist operators}
The family of minimal twist operators of the dual CFT$_5$ will be of particular importance to us. So, we introduce the notation $\O_\ell$ to denote the lowest dimensional primary operator with  spin $\ell$.\footnote{Twist of an operator with spin $\ell$ and dimension $\Delta$ is defined in the usual way $\tau=\Delta-\ell$.} In addition, we denote the anomalous dimension of $\O_\ell$ by $\gamma_\ell$. For the deformed generalized free CFT$_5$ dual to (\ref{ads}), $\O_\ell$ with even $\ell>1$ is always the double-trace operator $[\O^2]_{0,\ell}$. Whereas, for odd $\ell>1$, it is the triple-trace operator $[\O^3]_{0,\ell}$.\footnote{For large $\ell$, $[\O^3]_{0,\ell}$ can be degenerate. In that case, $\O_\ell$ (for odd $\ell$) represents the  $[\O^3]_{0,\ell}$ operator with the smallest anomalous dimension.} Consequently, anomalous dimensions $\gamma_\ell$ for $\ell>1$ are given by
\begin{align}
\gamma_\ell=&\gamma_{0,\ell}^{(2)}\ , \qquad \ell=\text{even}\ ,\nonumber \\
=&\gamma_{0,\ell}^{(3)}\ ,  \qquad \ell=\text{odd}\ .
\end{align}
The quantity $\gamma_\ell$ enjoys some nice properties. First of all, $\gamma_\ell$ asymptotes to zero
\be
\gamma_\ell \rightarrow 0 \qquad \text{as} \qquad \ell \rightarrow \infty\ .
\ee
Furthermore, $\gamma_\ell$ for even $\ell$ obeys the Nachtmann theorem which states that $\gamma_\ell$ is a monotonically increasing non-concave function of (even) $\ell\ge 2$ \cite{Nachtmann:1973mr,Komargodski:2012ek,Costa:2017twz,kundu}. The family of minimal twist operators with odd spins also obeys a generalized Nachtmann theorem which imposes lower bounds on $\gamma_\ell$ for odd $\ell$ \cite{kundu}.

\subsection{CFT Regge correlators}
We intend to map the $a$-theorem into a statement about anomalous dimensions $\gamma_2$ and $\gamma_3$ in the dual CFT$_5$. This can be achieved by studying various CFT four-point functions in the Regge limit. CFT four-point functions are highly constrained by conformal symmetries. In particular, a general Euclidean CFT four-point function of scalar primary operators can be written as
\be\label{G}
\langle O_1(x_1) O_2(x_2) O_3(x_3)O_4(x_4)\rangle =\frac{1}{(x_{12}^2)^{\frac{\Delta_1+\Delta_2}{2}}(x_{34}^2)^{\frac{\Delta_3+\Delta_4}{2}}}\left(\frac{x_{14}^2}{x_{24}^2} \right)^{\frac{\Delta_{21}}{2}}\left(\frac{x_{14}^2}{x_{13}^2} \right)^{\frac{\Delta_{34}}{2}} g(z,\bz)
\ee
where $x_{ij}^\mu=x_i^\mu-x_j^\mu$ and $\Delta_{ij}=\Delta_i-\Delta_j$. Conformal cross-ratios $z$ and $\bz$ are defined as follows
\begin{align}\label{cr}
z\bz=\frac{x_{12}^2 x_{34}^2}{x_{13}^2x_{24}^2}\ , \qquad (1-z)(1-\bz)=\frac{x_{14}^2x_{23}^2}{x_{13}^2x_{24}^2}\ .
\end{align}
Furthermore, the operator product expansion (OPE) enables us to write the function $g(z,\bz)$ as a sum over conformal blocks  
\be\label{schannel}
g(z,\bz)=\sum_p c_{12p}c_{34p}g^{\Delta_{12},\Delta_{34}}_{\Delta,\ell}(z,\bz)\ ,
\ee
where, $c_{ijk}$'s are OPE coefficients and the sum is over all primary operators of the theory. The conformal block expansion (\ref{schannel}) converges for Euclidean points $\bz=z^*$ with $|z|<1$ \cite{Mack:1976pa,Pappadopulo:2012jk}. 

\subsubsection*{Regge limit}
The CFT Regge limit  is an intrinsically Lorentzian limit of a Euclidean CFT four-point function. Lorentzian four-point functions can  be obtained as analytic continuations of the Euclidean correlator (\ref{G}). The analytic continuation is completely fixed by the ordering of operators in the Lorentzian correlator \cite{Hartman:2015lfa}. The CFT Regge limit is then defined by \cite{Brower:2006ea,Cornalba:2007fs,Cornalba:2008qf,Costa:2012cb}
\be\label{regge}
z,\bz\rightarrow 0\ , \qquad \text{with} \qquad \frac{\bz}{z}=\text{fixed}
\ee
of the Lorentzian four-point function. One way the Regge Lorentzian regime can be reached is by first rotating $z$ around the branch point of 1: $(1-z)\rightarrow (1-z)e^{-2\pi i}$, keeping $\bz$ fixed, and then taking the limit (\ref{regge}).

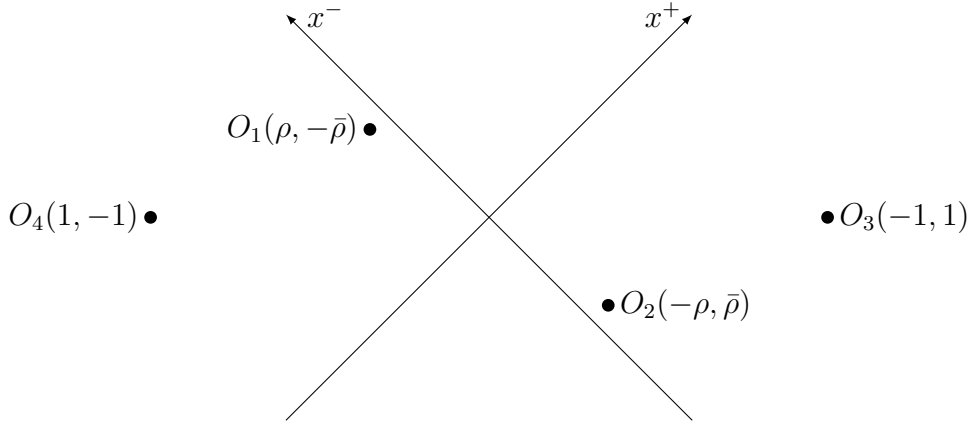
\begin{figure}
\begin{center}
\begin{tikzpicture}[baseline=-3pt, scale=1.80]
\begin{scope}[very thick,shift={(4,0)}]
\coordinate (v1) at (-1.5,-1.5) {};
\coordinate(v2) at (1.5,1.5) {};
\coordinate (v3) at (1.5,-1.5) {};
\coordinate(v4) at (-1.5,1.5) {};

\draw[thin,-latex]  (v1) -- (v2)node[left]{$x^+$};
\draw[thin,-latex]  (v3) -- (v4)node[right]{$\ x^-$};
\draw(-2.5,0)node[left]{ $\ O_4(1,-1)$};
\draw(2.5,0)node[right]{ $ O_3(-1,1)$};
\filldraw[black]  (-2.5,0) circle (1pt);
\filldraw[black]  (2.5,0) circle (1pt);
\coordinate(v5) at (0,0) {};
\def \fac {.6};
\filldraw[black]  (-0.88,0.65) circle (1 pt);
\filldraw[black]  (0.88,-0.65) circle (1pt);
\draw(-0.88,0.65)node[left]{ $ O_1(\rho,-\bar{\rho})$};
\draw(0.88,-0.65)node[right]{ $ O_2(-\rho,\bar{\rho})$};

\end{scope}
\end{tikzpicture}
\end{center}
\caption{\label{config} \small A Lorentzian four-point function where all points are restricted to a $2$d subspace. Null coordinates are defined as $x^{\pm}=x^0\pm x^1$, where $x^0$ is running upward. In the regime $0<\rho, \bar{\rho}<1$, this correlator is given by the Euclidean four-point function.}
\end{figure}

\begin{figure}[h!]
\centering
\includegraphics[scale=0.6]{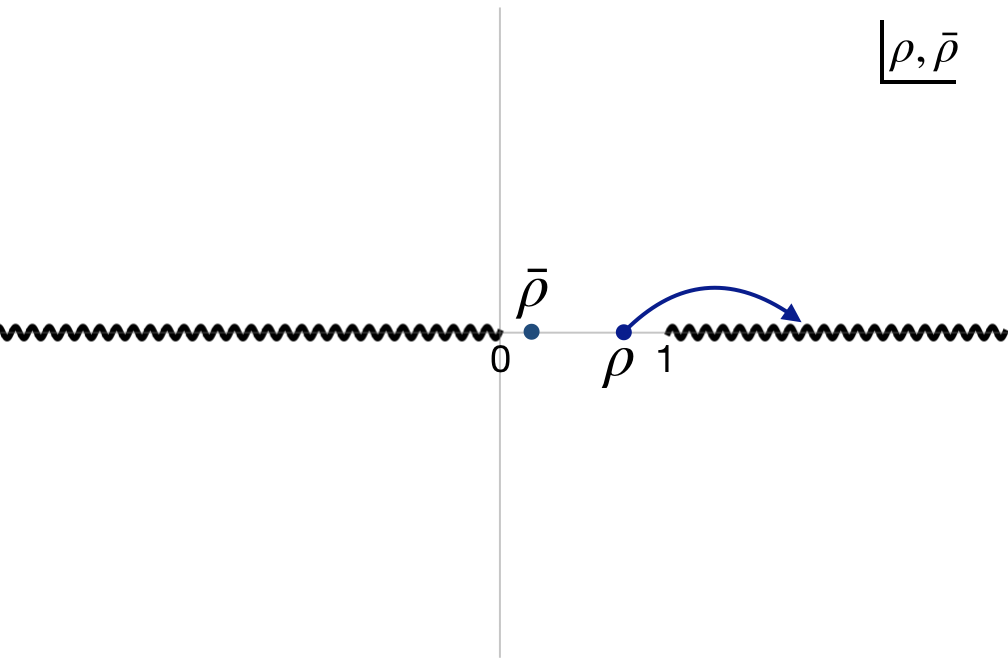}
\caption{ \label{dt1} \small Analytic structure of $G(\r,\br)$ -- branch cuts appear only when two operators become null separated. The Regge limit described above can be reached by analytically continuing $\r$ along the blue path.}
\end{figure}

One convenient way to describe the Regge limit is by starting with the Lorentzian correlator $G(\r,\br)=\langle O_1(x_1) O_2(x_2) O_3(x_3)  O_4(x_4)\rangle$ where all points are restricted to a $2$d subspace:
\begin{align}\label{points}
x_1=-x_2=(x^-=\rho,x^+=-\bar{\rho})\ ,  \qquad x_3=-x_4=(x^-=-1,x^+=1)\ ,
\end{align}
as shown in figure \ref{config}. Note that we are using null coordinates $x^{\pm}=x^0\pm x^1$. First, we restrict to the regime $0<\rho, \bar{\rho}<1$. In this regime, all points are space-like separated from each other and hence $G(\r,\br) $ is obtained trivially from (\ref{G})
\begin{align}\label{G2}
G(\r,\br) 
=\frac{2^{-\sum_i \Delta_i}}{(\rho \bar{\rho})^{\frac{\Delta_1+\Delta_2}{2}}}\left(\frac{(1-\rho ) \left(1-\bar{\rho }\right)}{(1+\rho ) \left(1+\bar{\rho }\right)}\right)^{\frac{\Delta_{21}+\Delta_{34}}{2}} \sum_p c_{12p}c_{34p}g^{\Delta_{12},\Delta_{34}}_{\Delta,\ell}(\rho,\bar{\rho})\ ,
\end{align}
where, $g^{\Delta_{12},\Delta_{34}}_{\Delta,\ell}(\rho,\bar{\rho})\equiv g^{\Delta_{12},\Delta_{34}}_{\Delta,\ell}(z(\rho),\bz(\bar{\rho}))$ with cross-ratios
\be\label{zzb}
z=\frac{4 \rho }{(1+\rho )^2}\ , \qquad \bz=\frac{4 \bar{\rho} }{(1+\bar{\rho} )^2}\ .
\ee
The s-channel expansion (\ref{G2}) converges for $|\r|,|\br|<1$. In general, the correlator $G(\r,\br)$ as a function of $\r$ and $\br$ is analytic in a larger domain as shown in figure \ref{dt1}.  Branch cuts appear only when two operators become null separated.

We now consider the Lorentzian corrrelator $\langle O_4(x_4) O_1(x_1) O_2(x_2) O_3(x_3)  \rangle$ where operators are ordered as written  with $\rho>1$. Note that operator pairs $O_4(x_4),O_1(x_1)$ and $O_2(x_2),O_3(x_3)$ are now time-like separated. This Lorentzian correlator is obtained from the Euclidean correlator by analytically continuing $\r$ along the path shown in figure \ref{dt1}. In terms of cross-ratios, this analytic continuation corresponds to $(1-z)\rightarrow (1-z)e^{-2\pi i}$ with $\bz$ fixed. The CFT Regge correlator is equivalently defined as the limit 
\be\label{regge2}
\rho\rightarrow \infty\ ,\quad \br\rightarrow 0\ , \qquad \text{with} \qquad \r \br=\text{fixed}
\ee
of the Lorentzian correlator $\langle O_4(x_4) O_1(x_1) O_2(x_2) O_3(x_3)  \rangle$.

The Regge limit of CFT correlators, being intrinsically Lorentzian, requires careful consideration. In Lorentzian CFT correlators, two operations -- analytic continuation and sum over conformal blocks in general may not commute. In fact, the Euclidean conformal block expansion (\ref{schannel}), as a series, does not converge when we take the Regge limit of individual conformal blocks. This problem was evaded in conformal Regge theory \cite{Costa:2012cb} by using the  Sommerfeld-Watson transform to resum the conformal block expansion (\ref{schannel}).
The conformal Regge theory exploits the fact that coefficients in the conformal block expansion (\ref{schannel}) are well defined analytic functions of spin \cite{Caron-Huot:2017vep}. This analyticity enables one to rewrite the conformal block expansion (\ref{schannel}) as a sum over Regge trajectories which is well behaved in the Regge limit \cite{Costa:2012cb}. To summarize, the Euclidean OPE of local operators is of limited use in the Regge limit. Instead, one should consider the contribution of  Regge trajectories to the OPE of local operators (see section 3.1 of \cite{Afkhami-Jeddi:2018apj} for details).

For our purpose, the family of minimal twist operators in deformed generalized free CFTs is of particular importance. Thus we can circumvent some of the intricacies of  conformal Regge theory by focussing on a very special limit $\frac{\bz}{z}\rightarrow 0$ of Regge correlators of the CFT$_5$ dual to (\ref{ads}). In this limit, parametric suppression of the  contributions from the spectral deformation  ensures that the sum over Regge limit of individual conformal blocks can still be trusted as an asymptotic series.  

\subsubsection*{Regge limit of conformal blocks}
In general, Regge conformal blocks of external scalar operators can be easily computed in any spacetime dimension by using the Regge OPE of \cite{Afkhami-Jeddi:2017rmx,Afkhami-Jeddi:2018own}. However, as explained earlier, we are only interested in the $\frac{\bz}{z}\rightarrow 0$ limit of Regge conformal blocks which leads to further simplification. For individual conformal blocks, the limit $\frac{\bz}{z}\rightarrow 0$ commutes with the Regge limit (\ref{regge}). This immediately implies that we can  start with lightcone conformal blocks and perform the appropriate analytic continuation to reach the Regge regime of interest. Lightcone conformal blocks are completely fixed by conformal symmetry. In particular,  conformal blocks  of external scalar primaries in the limit $\bz\rightarrow 0$  can be approximated in any spacetime dimension by \cite{Dolan:2000ut}
\be\label{lcblock}
g^{\Delta_{12},\Delta_{34}}_{\Delta,\ell}(z,\bz)\approx (-2)^{-\ell} \bz^{\frac{\Delta-\ell}{2}}z^{\frac{\Delta+\ell}{2}} {}_2F_1\left(\frac{\Delta+\ell-\Delta_{12}}{2},\frac{\Delta+\ell+\Delta_{34}}{2} ,\Delta+\ell,z\right)\ .
\ee
The lightcone conformal block has a branch cut along $z \in (1,\infty)$. One way the Regge  regime can be reached is by rotating $z$ around 1: $(1-z)\rightarrow (1-z)e^{-2\pi i}$. This analytic continuation can be implemented by using the identity (\ref{identity}). After using the identity (\ref{identity}), the Regge conformal blocks in the limit $\frac{\bz}{z}\rightarrow 0$ can be obtained from the lightcone block (\ref{lcblock})
\be\label{rgblock}
\tilde{g}^{\Delta_{12},\Delta_{34}}_{\Delta,\ell}(z,\bz)=i (-1)^{\ell}e^{-\frac{1}{2} i \pi  \left(\Delta _{12}-\Delta _{34}\right)}  \lambda^{\Delta_{12},\Delta_{34}}_{\Delta,\ell} \left(\frac{\bz}{z}\right)^{\frac{\Delta-\ell}{2}}\frac{1}{z^{\ell-1}}
\ee
where, $\lambda^{\Delta_{12},\Delta_{34}}_{\Delta,\ell}$ is a positive numerical coefficient given by
\be\label{lambda}
\lambda^{\Delta_{12},\Delta_{34}}_{\Delta,\ell}=\frac{2^{1-\ell}  \pi  \Gamma (\ell+\Delta -1) \Gamma (\ell+\Delta )}{\Gamma \left(\frac{ \ell+\Delta -\Delta _{12}}{2}\right) \Gamma \left(\frac{\ell+\Delta +\Delta _{12}}{2}\right) \Gamma \left(\frac{\ell+\Delta -\Delta _{34}}{2} \right) \Gamma \left(\frac{\ell+\Delta +\Delta _{34}}{2} \right)}\ .
\ee

\subsection{Conformal bootstrap}
The purpose of this paper is to make the CFT-based description of RG flows, as shown in figure \ref{intro}, explicit by relating $\hat{b}$ and $\Delta a$ to CFT$_5$ data. First, we consider the generalized free CFT in 5d which is dual to the AdS theory (\ref{ads}) with $\hat{b}=\Delta a=0$. The physical dilaton is dual to the operator $\O$. The Euclidean four-point function $\langle \O(x_1)\O(x_2)\O(x_3)\O(x_4) \rangle$  for $0<\r,\br<1$ is given by \footnote{Points $x_1,x_2,x_3$ and $x_4$ are given by (\ref{points}).}
\begin{align}\label{2pt}
G_4(\r,\br)=\frac{c_\O^2}{(16 \r\br)^{\Delta_\O}}+\frac{c_\O^2}{((1-\r)(1-\br))^{2\Delta_\O}}+\frac{c_\O^2}{((1+\r)(1+\br))^{2\Delta_\O}}
\end{align}
which is trivially crossing symmetric. Note that the operator $\O$ is not canonically normalized and $c_\O>0$ is the coefficient of the $\langle \O\O\rangle$ two-point function. 

It should be possible to express the above four-point function as a sum over conformal blocks. In particular, following \cite{Heemskerk:2009pn} we can write (\ref{2pt}) as an s-channel sum over $[\O^2]_{n,\ell}$ exchanges 
\be\label{schannel}
G_4(\r,\br)=\frac{1}{(16\r\br)^{\Delta_\O}}\left(c_\O^2+\sum_{n,\ell=0}^\infty c(n,\ell)^2\ g_{\Delta^{(2)}_0(n,\ell),\ell}^{0,0}(\r,\br)\right)\ ,
\ee
where, $\Delta^{(2)}_0(n,\ell)=2\Delta_\O+2n+\ell$ is the dimension of $[\O^2]_{n,\ell}$ as given in (\ref{delta}) without the anomalous dimension part. The OPE coefficients $c(n,\ell)$ are completely known from \cite{Heemskerk:2009pn}, however, only useful information that we need is that the OPE coefficients are real because of unitarity.

\subsubsection{Mixed correlators}
 We now make a little detour. In order to gain some more insight, we consider a generalized free CFT with two operators $\O_1$ and $\O_2$ that are dual to two free (massive or massless) fields $\phi_1$ and $\phi_2$ in AdS. Now, we can study a mixed correlator in the Euclidean regime  $0<\r,\br<1$ for the kinematics (\ref{points})
 \be\label{mixed}
 \tilde{G}_4(\r,\br)=\langle \O_2(x_1)\O_1(x_2)\O_1(x_3)\O_2(x_4) \rangle=\frac{c_{\O_1}c_{\O_2}}{((1-\r)(1-\br))^{\Delta_{\O_1}+\Delta_{\O_2}}}\ .
 \ee
Again we can express the t-channel identity exchange as an s-channel sum over double-trace operators $[\O_1\O_2]_{n,\ell}$
\be
\tilde{G}_4(\r,\br)=\frac{1}{(16\rho \bar{\rho})^{\frac{\Delta_{\O_1}+\Delta_{\O_2}}{2}}}\left(\frac{(1-\rho ) \left(1-\bar{\rho }\right)}{(1+\rho ) \left(1+\bar{\rho }\right)}\right)^{\Delta_{\O_1}-\Delta_{\O_2}}\sum_{n,\ell=0}^\infty (-1)^{\ell}\ \tilde{c}(n,\ell)^2\ g_{\Delta^{(1,2)}_0(n,\ell),\ell}^{\Delta_{21},-\Delta_{21}}(\r,\br)\ ,
\ee
where, $\Delta^{(1,2)}_0(n,\ell)=\Delta_{\O_1}+\Delta_{\O_2}+2n+\ell$ is the dimension of operators $[\O_1\O_2]_{n,\ell}$. The s-channel OPE coefficients $\tilde{c}(n,\ell)$ are uniquely determined by the t-channel identity exchange. 

\subsubsection{Correlators of double-trace operators}
We are now ready to study mixed correlators in the generalized free CFT in 5d which is dual to the AdS$_6$ theory (\ref{ads}). The correlator that will be of significant importance is very similar to the correlator (\ref{mixed}) with one key difference -- the operator $\O_2=\O^2$ is now a double-trace operator
\be\label{double}
\O^2 \equiv [\O^2]_{n=0,\ell=0}= \lim_{x'\rightarrow x}\frac{1}{\sqrt{2}} \O(x')\O(x)\ .
\ee
The mixed four-point function $G_{mixed}(\r,\br)=\langle \O^2(x_1)\O(x_2)\O(x_3)\O^2(x_4) \rangle$ even for the generalized free theory ($\hat{b}=\Delta a=0$) appears to be more complicated. In the kinematics (\ref{points}),  $G_{mixed}(\r,\br)$ is given by
\be\label{mixed_G}
((1-\r)(1-\br))^{\Delta_\O}G_{mixed}(\r,\br)=\frac{2c_\O^3}{(16\r\br)^{\Delta_\O}}+\frac{c_\O^3}{((1-\r)(1-\br))^{2\Delta_\O}}+\frac{2c_\O^3}{((1+\r)(1+\br))^{2\Delta_\O}}\ .
\ee
The above discussion about four-point functions applies to correlators of double-trace operators as well. So, the above correlator  can also be written as a similar s-channel expansion 
\begin{align}\label{mixed_s}
G_{mixed}(\r,\br)=&\frac{1}{(16\r\br)^{\frac{3\Delta_{\O}}{2}}}\left(\frac{(1+\rho ) \left(1+\bar{\rho }\right)}{(1-\rho ) \left(1-\bar{\rho }\right)}\right)^{\Delta_{\O}}\nonumber\\
&\times 
\left(c_{\O\O\O^2}^2\ g^{\Delta_\O,-\Delta_\O}_{\Delta_\O,0}(\r,\br)+\sum_{[\O^3]_{n,\ell}} (-1)^{\ell}\ \tilde{c}(n,\ell)^2\ g_{\Delta^{(3)}_0(n,\ell),\ell}^{\Delta_{\O},-\Delta_{\O}}(\r,\br)\right)
\end{align}
where, $\Delta^{(3)}_0(n,\ell)=3\Delta_\O+2n+\ell$ is the dimension of $[\O^3]_{n,\ell}$ as given in (\ref{delta}) without the anomalous dimension.  The first term in (\ref{mixed_s}) corresponds to the exchange of $\O$ with the OPE coefficient $c_{\O\O\O^2}=\sqrt{2} c_\O^{3/2}$.\footnote{Note that the OPE coefficients are appropriately normalized.} Hence, this term is exactly the first term of (\ref{mixed_G}). On the other hand, the s-channel sum over triple-trace operators $[\O^3]_{n,\ell}$ reproduces the remaining two terms of (\ref{mixed_G}).

\subsection{Anomalous dimensions}
Let us now turn on interactions $\mathcal{L}_{int}$ in (\ref{ads}). Four-point functions of the dual CFT$_5$ are now perturbative solutions to crossing symmetry -- each interaction of the AdS$_6$ scalar field $\phi$ leads to a specific contribution to the anomalous dimensions $\gamma_{n,\ell}^{(N)}$ of multi-trace operator $[\O^N]_{n,\ell}$. The whole purpose of this section is to build up a simple framework that enables the identification of the contributions of different AdS$_6$ interactions to $\gamma_{n,\ell}^{(N)}$.

First we consider the four-point function (\ref{schannel}). In the presence of $\hat{b}$ and $\Delta a$, this four-point function receives corrections $\delta G_4(\r,\br)$ which can be computed using the conventional AdS perturbation theory. On the CFT$_5$ side, $\delta G_4(\r,\br)$ originates from anomalous dimensions $\gamma_{n,\ell}^{(2)}$ of double-trace operators  and corrections of their OPE coefficients $c(n,\ell)+\delta c(n,\ell)$. In particular, using (\ref{rgblock}) and (\ref{zzb}) we obtain the leading contribution of $\gamma_{n,\ell}^{(2)}$ and $\delta c(n,\ell)$ to the four-point function in the Regge limit (\ref{regge2}) followed by the limit $|\gamma_{n,\ell}^{(2)}|\ll\r\br\ll 1 $
\be\label{formula1}
\delta G_4(\r,\br)|_{\gamma_{n,\ell}^{(2)},\delta c(n,\ell)}=i \frac{c(n,\ell)^2 \gamma^{(2)}_{n,\ell}}{2^{4\Delta_\O+2\ell-1}} \lambda^{0,0}_{2\Delta_\O+2n+\ell,\ell} \left(\r\br\right)^{n}\ \log \left(\r\br\right)\r^{\ell-1}+\cdots\
\ee
with even $\ell$ and $\Delta_\O=5$, where dots represent subleading corrections without the $\log(\r\br)$.  The numerical factor $\lambda$ is given in (\ref{lambda}).

Before proceeding further, let us offer some more comments. First, note that $\delta c(n,\ell)$ does not contribute to the leading $\log(\r\br)$ term in (\ref{formula1}). Moreover, the expression (\ref{formula1}) implies that for a fixed $\ell$, the dominant contribution  in the limit $|\gamma_{n,\ell}^{(2)}|\ll\r\br\ll 1 $ always comes from $[\O^2]_{0,\ell}\equiv \O_\ell$ which, as stated earlier, is of importance to us. 

Likewise, we can obtain the leading contribution of $\gamma_{n,\ell}^{(3)}$ and $\delta \tilde{c}(n,\ell)$ to the four-point function $G_{mixed}(\r,\br)$ of the double-trace operator (\ref{mixed_G}). In the Regge limit followed by the limit $|\gamma_{n,\ell}^{(3)}|\ll \r\br\ll 1 $ we can now write
\be\label{formula2}
\delta G_{mixed}(\r,\br)|_{\gamma_{n,\ell}^{(3)},\delta \tilde{c}(n,\ell)}=i \frac{\tilde{c}(n,\ell)^2 \gamma^{(3)}_{n,\ell}}{2^{6\Delta_\O+2\ell-1}} \lambda^{\Delta_\O,-\Delta_\O}_{3\Delta_\O+2n+\ell,\ell}\left(\r\br\right)^{n}\ \log \left(\r\br\right)\r^{\ell-1}+\cdots\
\ee
where $\Delta_\O=5$ and dots again represent subleading corrections without the $\log(\r\br)$. This result is qualitatively very similar to (\ref{formula1}), for example  at fixed $\ell$, the dominant contribution to $\delta G_{mixed}(\r,\br)$ also comes from the lowest twist triple-trace operator $[\O^3]_{0,\ell}$. However, there is one difference -- unlike $(\ref{formula1})$ both even and odd spin operators contribute to $\delta G_{mixed}(\r,\br)$.

In the rest of the paper, equations (\ref{formula1}) and (\ref{formula2}) will play important roles. So, we should examine them more closely.  It is easy to see that  contributions of individual operators in (\ref{formula1}) and (\ref{formula2}) become increasingly singular with increasing spin implying that in general we should not trust the Regge limit of individual conformal blocks. On the other hand, it was argued in \cite{Hartman:2015lfa} that analytically continued s-channel conformal blocks still can be trusted in the lightcone limit. The same argument applies here as well provided $\gamma_{n,\ell}^{(N)},\delta \tilde{c}(n,\ell)$, and $\delta c(n,\ell)$ are parametrically suppressed with increasing $\ell$. This is precisely the case for the CFT$_5$ dual to (\ref{ads}) which is after all an effective field theory in AdS. Consequently,  $\gamma_{n,\ell}^{(N)},\delta \tilde{c}(n,\ell)$, and $\delta c(n,\ell)$ are suppressed by increasing powers of $1/\Delta_f$ as we increase spin implying that equations (\ref{formula1}) and (\ref{formula2}) are still reliable for the dual CFT$_5$.\footnote{Note that the gap is defined as $\Delta_f=R_{\text{AdS}}f\gg 1$.}

\addtocontents{toc}{\protect\setcounter{tocdepth}{1}}
\section{The $a$-theorem and anomalous dimensions}\label{section_atheorem}
\addtocontents{toc}{\protect\setcounter{tocdepth}{-1}}
We are now in a position to relate $\hat{b}$ and $\Delta a$ to anomalous dimensions $\gamma_2\equiv \gamma_{0,2}^{(2)}$ and $\gamma_3\equiv \gamma_{0,3}^{(3)}$ by studying Lorentzian four-point functions for the CFT$_5$ dual to the effective field theory (\ref{ads}). It  will be discussed in length below.

Lorentzian correlators are analytic continuations of Euclidean correlators. So, it is equivalent to study the CFT$_5$ in the Euclidean signature which is now dual to the 6d bulk Euclidean theory
\be\label{eads}
S^E_{eff}[\phi]=\int d^6x \sqrt{g_{EAdS}}\left(\frac{1}{2}g^{\mu\nu}_{EAdS}\partial_\mu \phi \partial_\nu \phi- \mathcal{L}_{int}\right)\ .
\ee
In the above action, all derivatives in $\mathcal{L}_{int}$ now are taken using the Euclidean AdS metric. The equation of motion for the dilaton field $\phi$ is given by
\begin{align}
\Box \phi= -\frac{\delta \mathcal{L}_{int}}{\delta \phi}\equiv&-\frac{\hat{b}}{2f^4}\left(2\phi \Box^2 \phi+ \Box^2 \phi^2 \right)-\frac{\hat{b}}{4f^6}\left( 3\phi^2 \Box^2 \phi +  \Box^2 \phi^3+\phi\Box^2 \phi^2 \right)\nonumber\\
&-\frac{\hat{b}}{32f^8}\left(20 \phi^3 \Box^2 \phi +5  \Box^2 \phi^4+6\phi^2 \Box^2 \phi^2+4\phi \Box^2 \phi^3\right)\nonumber\\
&+\cdots\ ,
\end{align}
where, dots represent terms that contribute to four-point and six-point functions only at subleading orders. Under the boundary condition $\phi(x,\z\rightarrow 0)=\Phi(x)$, the above equation of motion has the formal solution
\be\label{bulk}
\phi(\z,x)=\int d^5 x' K(\mathfrak{z},x;x')\Phi(x')- \int d^{5}x' d\z' \sqrt{g_{EAdS}(\z')} G(\z,x;\z',x') \frac{\delta  \mathcal{L}_{int}}{\delta \phi}(\z',x')
\ee
where, $\mathfrak{ z}$ is the bulk direction in AdS and $x\in \mathbb{R}^5$ (the metric is given by (\ref{metric})).  In the above expression, $K(\mathfrak{z},x;x')$ is the bulk-to-boundary propagator, whereas $G(\z,x;\z',x')$ is the bulk-to-bulk propagator for the dilaton field which are transcribed in appendix \ref{holography}.\footnote{For convenience, we are using $R_{\text{AdS}} = 1$. We will restore $R_{\text{AdS}}$ by dimensional analysis whenever necessary.}

\subsection{On-shell action}
The asymptotic value $\Phi(x)$ of the dilaton field  acts as the source for the CFT primary operator $\O(x)$. In principle, any tree-level correlator of $\O$ can be computed from the bulk on-shell action $S_{on-shell}^E[\Phi]$ which determines the CFT partition function $Z[\Phi]=\exp\(-S_{on-shell}^E[\Phi]\)$ \cite{Maldacena:1997re, Witten:1998qj,Gubser:1998bc}. Thus, we should start by examining the bulk on-shell action more closely.

In general, the bulk on-shell action diverges as we take the UV cut-off $\epsilon\rightarrow 0$. This divergence can be removed by adding a counter-term on the boundary $\z=\epsilon$. However, for massless fields this diverging piece vanishes exactly and hence  boundary terms are not required to make the on-shell action finite. Of course, there still can be other divergences coming from loops in the bulk. These are standard QFT divergences which can be removed by adding bulk counter-terms.

The total on-shell action for the Euclidean theory (\ref{eads}) can be written in a compact form 
\begin{align}\label{tyrion}
S_{on-shell}^E&=-\frac{80}{\pi^3}\int_{\z=\epsilon}d^5 x_1 d^5 x_2  \frac{\Phi(x_1)\Phi(x_2)}{|x_1-x_2|^{10}}-\int d^{5}x\ d\z \sqrt{g_{EAdS}(\z)} \mathcal{L}_{int}(z,x)\\
&-\frac{1}{2}\int d^{5}xd\z \sqrt{g_{EAdS}(\z)}\int d^{5}x' d\z' \sqrt{g_{EAdS}(\z')} G(\z,x;\z',x')\frac{\delta \mathcal{L}_{int}}{\delta \phi}(\z',x')\frac{\delta \mathcal{L}_{int}}{\delta \phi}(\z,x)\ ,\nonumber
\end{align}
where, $\z=\epsilon$ is the UV cut-off and the bulk field $\phi$ should be understood to be the solution (\ref{bulk}). This form of the on-shell action is useful for performing a systematic perturbative expansion. This is exactly what we need to do since the bulk theory only makes sense perturbatively.

As a warm up, we start with the two-point function $\langle\O\O\rangle$. The form of the on-shell action (\ref{tyrion}) makes it particularly easy to read off the two-point function 
\be
\langle \O(x_1) \O(x_2)\rangle =\(\frac{160}{\pi^3}\)\frac{1}{|x_1-x_2|^{10}}\ 
\ee
implying $\Delta_\O=5$ and $c_\O=\frac{160}{\pi^3}$. For higher point functions, one needs to insert the solution (\ref{bulk}) in (\ref{tyrion}) and perform a perturbative expansion in $1/f$. Thus, from here on calculations are going to be more involved.

\subsection{Four-point function and $\gamma_2$}\label{spin2}
Consider the contribution of the first term of equation (\ref{tyrion}) to the four-point function of operator $\O$. Clearly, this contribution reproduces the four-point function (\ref{2pt}) of the generalized free theory. The leading correction to the four-point function comes from the second term of (\ref{tyrion}). In particular, using the explicit form of $\mathcal{L}_{int}$ (\ref{int}), we can obtain the leading interacting term of the on-shell action with four dilatons
\be\label{ma1}
S^{(\phi^4)}_{on-shell}=-\frac{\hat{b}}{16 f^6}\int d^{5}x\ d\z \sqrt{g_{EAdS}(\z)}  \phi^2 \Box^2 \phi^2 +\cdots\ ,
\ee
where dots represent terms that are subleading. This on-shell action can be expanded by using the bulk solution (\ref{bulk}) and at the leading order in perturbation theory, using the identity (\ref{identity1}) we obtain
\begin{align}\label{ma2}
S^{(\phi^4)}_{on-shell}=-\frac{\hat{b}}{4 f^6}\left( \frac{32}{\pi ^3}\right)^4 5^4 \int_{\Phi^4}\int_{AdS}&\left(\tilde{K}_{5}(y_3)\tilde{K}_{5}(y_4)-2 y_{34}^2 \tilde{K}_{6}(y_3)\tilde{K}_{6}(y_4) \right)\nonumber\\
&\times \left(\tilde{K}_{5}(y_1)\tilde{K}_{5}(y_2)-2 y_{12}^2 \tilde{K}_{6}(y_1)\tilde{K}_{6}(y_2) \right)\ ,
\end{align}
where, we have introduced
\be\label{notation}
\int_{AdS}\equiv \int d^{5}x\ d\z \sqrt{g_{EAdS}(\z)}\ , \qquad \int_{\Phi^N} \equiv \prod_{i=1}^N\int  \Phi(y_i)d^dy_i
\ee
and the reduced bulk-to-boundary propagator $\tilde{K}_{\Delta}(y)$ (see equation (\ref{reduced})) to lighten the notation.

  Now consider the Lorentzian correlator $G_4(\r,\br)=\langle \O(x_4) \O(x_1) \O(x_2) \O(x_3)  \rangle$ in  the kinematics (\ref{points}), where operators are ordered as written. 
At the tree-level, the correlator  $G_4(\r,\br)$   is schematically given by the Witten diagram 
\be\label{diagram1}
\includegraphics[scale=0.5]{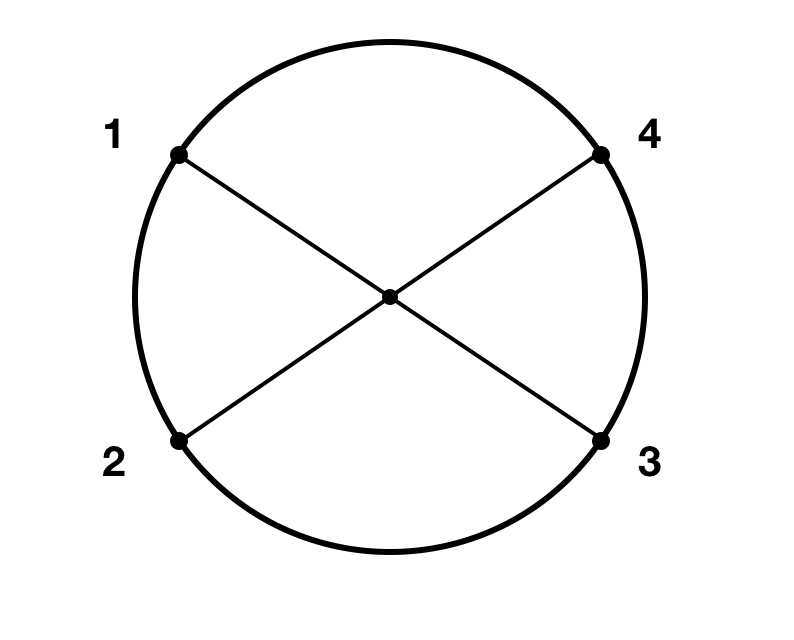}\ .
\ee
We wish to determine the  contribution of (\ref{ma1}) to $G_4(\r,\br)$ in the Regge limit (\ref{regge2}). It is a straightforward exercise to compute the  leading Regge contribution from the on-shell action (\ref{ma2})
\be
\delta G_4(\r,\br)= \frac{16\hat{b}}{ \Delta_f^6}\left( \frac{32}{\pi ^3}\right)^4 5^4 \r^2 D_{6666}(\r,\br)
\ee
where the $D$-function is defined in (\ref{defineD}) and $\Delta_f=R_{\text{AdS}} f\gg 1$. This $D$-function can be calculated exactly using the integral expression (\ref{calcD}). However, we are only interested in a specific limit of the Regge correlator: $\r\br\rightarrow 0$. In this limit, the above expression can be simplified further by using (\ref{finalD})
\be\label{finalg4}
\delta G_4(\r,\br)\approx -i \frac{4\hat{b}}{ \Delta_f^6}  \frac{5^4}{\pi^{17/2}}\frac{\Gamma(11)\Gamma\(\frac{19}{2}\)}{\Gamma(6)^4}\r \log\(\r\br\)\ .
\ee
Now, we must compare this result with (\ref{formula1}). Clearly, this contribution can only come from anomalous dimension of the operator $[\O^2]_{0,\ell=2}$. This enables us to relate the anomalous dimension $\gamma_2$ with $\hat{b}$
\begin{bBox}
\be\label{gamma2}
\gamma_2=- \(\frac{51}{8 \pi ^3}\) \frac{\hat{b}}{ \Delta_f^6}
\ee
\end{bBox}
where, we have used equation (\ref{lambda}) and the generalized free field value of $c(0,2)^2=\frac{300}{11}c_0^2$. It is not a surprise that $\gamma_2$ is related to $\hat{b}$. After all, in \cite{Heemskerk:2009pn} it has been established conclusively that a bulk interaction $\phi^2\Box^{2k}\phi^2$ corresponds to  anomalous dimensions of $[\O^2]_{n,k}$ double-trace operators at the tree level. Nonetheless, this exercise highlights  the core of our argument that we will apply to mixed correlators. Before we proceed to more involved mixed correlators, let us make some comments that will be useful.

\subsubsection*{Subleading corrections}
What does the second part of the 4-point dilation interaction (\ref{int4}) correspond to in the dual CFT$_5$? Clearly, the term $\Delta a\ \phi^2\Box^3\phi^2$ cannot contribute to anomalous dimensions of spin-3 double-trace operators since there are no spin-3 double-trace operators. However, it does contribute to anomalous dimensions of spin-2 double-trace operators $[\O^2]_{n,2}$ but at a subleading order in $1/\Delta_f$. This can be seen by repeating the preceding analysis for this six-derivative interaction. The final result will have the same functional behavior as (\ref{finalg4}) but with the pre-factor $\frac{\Delta a}{\Delta_f^8}$. What distinguishes the six derivative interaction from the four derivative interaction is that the resulting anomalous dimensions $\gamma_{n,\ell=2}^{(2)}$ have different asymptotic behaviors for large $n$ \cite{Heemskerk:2009pn}. Furthermore, the dilaton 3-point interaction (\ref{int3}) also contributes to anomalous dimensions of spin-2 double-trace operators $[\O^2]_{n,2}$ at the order $1/\Delta_f^8$.

Finally, we should discuss subleading corrections from loop contributions. Of course, we expect to get UV divergences from loop diagrams. However, at any order in the perturbation theory, these divergences can be removed by adding a finite number of contact interactions. The point we wish to establish is that  loop diagrams, however complicated, can only contribute to $G_4(\r,\br)$ in a very specific way in the Regge limit. Let us demonstrate this for loop diagrams at the order $\frac{1}{f^{12}}$.

Crossing symmetry requires that one-loop diagrams must contribute to anomalous dimensions for all double-trace operators $[\O^2]_{n,\ell}$ with any $n$ and (even) $\ell$ \cite{Aharony:2016dwx}. Some of these anomalous dimensions are UV divergent. In particular, for loop diagrams at the order $\frac{1}{f^{12}}$,  anomalous dimensions $\gamma^{(2)}_{n,\ell\le4}$ can diverge. These divergences can be removed by adding counter-terms $\frac{1}{f^{12}}\phi^2 \Box^{2k}\phi^2$ with $k=0,1,\cdots,4$. Needless to say that there still can be a finite $\gamma^{(2)}_{n,\ell\le4}\sim \frac{1}{\Delta_f^{12}}$ remaining after this subtraction.  On the other hand, simple power counting implies that $\gamma^{(2)}_{n,\ell}$ is finite for $\ell>4$. Hence, we can write
\be
\gamma^{(2)}_{n,\ell}\approx \gamma^{(2)}_{n,\ell}|_{\text{contact}}+\gamma^{(2)}_{n,\ell}|_{\text{1-loop}}\sim \frac{1}{\Delta_f^{2+2\ell}}+\frac{1}{\Delta_f^{12}}\,
\ee
where, $\gamma^{(2)}_{n,\ell}|_{\text{contact}}$ comes from contact interactions of the original dilaton effective theory. Clearly, the 1-loop contributions dominate for $\ell\ge 6$. At first sight, this appears to be in contradiction with the analytic structure of $G_4(\r,\br)$ which implies that if the part of $G_4(\r,\br)$ that grows with $\r$ in the Regge limit admits an expansion in $\r$
\be\label{con_anc}
G_4(\r,\br) \sim i \sum_{L=2,4,6,\cdots}  c_L \r^{L-1}
\ee
then coefficients $c_L$ must obey the following properties for any $L$ \cite{kundu}: (i) $c_L\ge 0$, (ii) $\frac{c_{L+2}}{c_L}$ must be parametrically suppressed, (iii) $c_{L+2}^2\le c_L c_{L+4}$ . Obviously, if we first take the Regge limit of individual conformal blocks, condition (ii) is in tension with (\ref{formula1}) for $L\ge 6$. This has led us to make the following two important conclusions. 

The first is that at the 1-loop level, summing over an infinite set of conformal blocks and taking the Regge limit -- these two operations do not commute. The second conclusion is that when we sum over conformal blocks and then take the Regge, as we should do, the full 1-loop contribution can only be consistent with (\ref{con_anc}) if and only if it does not grow faster than
\be\label{loop}
G^{\text{1-loop}}_4(\r,\br) \sim i \frac{\r^3}{\Delta_f^{12}}+\O\(\frac{1}{\Delta_f^{14}}\)
\ee
in the Regge limit. This fact will be important in the discussion of mixed correlators. Note that this is actually a conservative bound and crossing symmetry may impose a stronger restriction.

\subsection{Positivity}
We begin with a discussion about a general positivity condition that applies to any unitary RG flow.  Let us consider an RG flow in $d$ dimensions that connects two conformal fixed points $\CFTUV$ and $\CFTIR$. We have argued that any such flow can be equivalently described by spectral deformations of a generalized free CFT in $d-1$ dimensions, as summarized in figure \ref{intro}. In this description, the starting point of the RG flow corresponds to the generalized free CFT$_{d-1}$ in which we consider the Lorentzian correlator $G_4(\r,\br)=\langle \O(x_4) \O(x_1) \O(x_2) \O(x_3)  \rangle$, where operators are ordered as written with $\r>1$ and $0<\br<1$.  Clearly, this four-point function is the analytic continuation of the Euclidean correlator (\ref{2pt}) with $\Delta_\O=d-1$. Now, the RG flow deforms the dual CFT$_{d-1}$  and hence at the end of the RG flow the  four-point function becomes $G_4(\r,\br)+\delta G_4(\r,\br)$. For every unitary RG flow, this deformed correlator  must obey Rindler positivity. In particular, in the Regge limit (\ref{regge2}) the argument of \cite{Hartman:2016lgu} can be easily extended to conclude 
\be
-\mbox{Re}\ \delta G_4(\r,\br) \ge 0\ .
\ee
For unitary RG flows in even spacetime dimensions, this positivity condition is of particular importance.

\subsubsection*{RG flows in 6d}
Classical causality argument of \cite{Adams:2006sv} suggests that $\hat{b}$ is non-negative. However, the dispersive sum-rule (\ref{sumrule}), as pointed out in \cite{Elvang:2012st}, requires an additional assumption about the asymptotic behavior of the 4-point on-shell scattering amplitude $\mathcal{A}(s,t)$ of dilaton. In contrast, the negativity of $\gamma_2$ follows directly from the CFT Nachtmann theorem \cite{Komargodski:2012ek,Costa:2017twz,kundu}, as well as from causality \cite{Hartman:2015lfa}. Moreover, the analyticity of CFT correlators in Lorentzian signature, as explained in  \cite{Hartman:2015lfa,Hartman:2016lgu}, allows us to write a CFT$_5$ sum-rule for $\hat{b}$
\be\label{CFTsumrule}
\hat{b}=\frac{\Delta_f^6 \pi^{15/2}\Gamma(6)^4}{2(5)^4 \Gamma(11)\Gamma\(\frac{19}{2}\)}\lim_{\eta\rightarrow 0}\lim_{x\rightarrow 0}\frac{1}{\log \eta} \int^x_{0} d\sigma\ \mbox{Re}\ \delta G_4\(\r=\frac{1}{\sigma},\br=\eta \sigma\)\ge 0\ ,
\ee
where we have utilized the fact that $ \delta G_4(\sigma)= \delta G_4(-\sigma)$ on the real line. This sum-rule  does not make any assumptions about the dual CFT$_5$ beyond the usual Euclidean axioms. This suggests that some properties of effective field theory, perhaps surprisingly, are more transparent in AdS.

We should emphasize that conceptually $\hat{b}\ge0$ is a non-trivial dynamical inequality. The parameter $\hat{b}$ is a complicated quantity that depends both on $\CFTUV$, $\CFTIR$ and the details of the RG flow. So, on one hand it is indeed surprising that it obeys a positivity condition, but  on the other the dynamical nature of the inequality suggests that this positivity condition is of little practical importance. Nonetheless, it surely makes us wonder whether a similar positivity condition for $\Delta a$ in 6d follows from the requirement that the dual CFT$_5$ must be well behaved. 

\subsubsection*{RG flows in 4d}
We end this discussion with some comments about unitary RG flows in 4d. In this case, $\Delta a$ appears in the dilaton effective action as the coefficient of the four derivative term $\phi^2\Box^2\phi^2$. The analysis of this section can be repeated almost exactly to obtain the Regge correlator $ \delta G_4\sim -i \frac{\Delta a}{\Delta_f^4}\r \log(\r\br)$. This immediately implies that $\Delta a$ in 4d also obeys the CFT sum-rule (\ref{CFTsumrule}), however,  with a different factor in front. Furthermore, similar to \cite{Komargodski:2011vj}, we can also construct a CFT$_3$ quantity that decreases monotonically along the flow  
\be
a(\mu)=a_{UV}-\tilde{\Delta}_f^4\lim_{\eta\rightarrow 0}\lim_{x\rightarrow 0}\frac{1}{\log \eta} \int^x_{\mu x} d\sigma\ \mbox{Re}\ \delta G_4\(\eta, \sigma\)\ .
\ee
This quantity interpolates between $a(\mu\rightarrow 0)=a_{IR}$ and $a(\mu\rightarrow 1)=a_{UV}$.\footnote{Note that the actual value of $\Delta_f=R_{\text{AdS}}f$ has no significance. So, we can always redefine $\Delta_f$ by absorbing some positive numerical factors, which we will denote by the symbol $\tilde{\Delta}_f$.}

\subsection{Six-point function and $\gamma_3$}
Clearly the preceding analysis can be extended to the mixed correlator $G_{mixed}(\r,\br)=\langle \O^2(x_4)  \O^2(x_1)\O(x_2)\O(x_3)\rangle$ which involves computation of six-point functions.\footnote{The double-trace operator $\O^2$ is defined in (\ref{double}).} The task of calculating the six-point function may seem challenging,  however, it simplifies once we figure out exactly what we are looking for. We begin by asking what CFT quantity receives leading contribution from the second part of the six-point interaction (\ref{int6}). Of course, anomalous dimensions of spin-2 triple-trace operators $[\O^3]_{n,2}$ cannot be the answer, since the dominant contribution to $\gamma^{(3)}_{n,2}$ always comes from four-derivative interactions. So, we consider spin-3 triple-trace operators $[\O^3]_{n,3}$ which do exist. The above discussion has tempted us to expect that the six-derivative part of the six-point interaction can contribute to anomalous dimensions of $[\O^3]_{n,3}$. This possibility is particularly promising because the four-derivative part of the six-point interaction is not expected to contribute to $\gamma^{(3)}_{n,3}$ at the tree-level. In the rest of this section, we will show that both of these expectations are indeed true.

We already know  from equation (\ref{formula2}) how the mixed correlators $G_{mixed}(\r,\br)$ must behave in the Regge limit followed by the limit $\r\br\rightarrow 0$ if operators $[\O^3]_{n,3}$ acquire anomalous dimensions. The unique leading contribution is completely fixed by the anomalous dimension $\gamma_3$ of the lowest twist spin-3 triple-trace operators $[\O^3]_{0,3}$
\be\label{growth}
\delta G_{mixed}(\r,\br)\sim i \gamma_3 \log \left(\r\br\right)\r^{2}\ .
\ee
Thus it is sufficient for us to only compute the part of $G_{mixed}(\r,\br)$ in the Regge limit followed by the limit $|\gamma_3|\ll \r\br\ll 1$ that has the above behavior. Of course, we must not assume that only $\Delta a$ can contribute to $\gamma_3$. We need to consider all possible contributions to $\gamma_3$ up to order $\frac{1}{\Delta_f^{12}}$. Moreover, we should also remember that some part of $\gamma_3$ may not contribute to a growth like (\ref{growth}) in the Regge limit. Thus, let us write $\gamma_3$ as
\be
\gamma_3=\gamma_3^{\text{Regge}}+\gamma_3^{\text{other}}\ ,
\ee
where $\gamma_3^{\text{other}}$ is the part of $\gamma_3$ that does not contribute to the Regge growth (\ref{growth}). This part of $\gamma_3$ is difficult to compute, however, we can still derive general results about $\gamma_3^{\text{other}}$.

Let us now compute that relevant part of the Lorentzian  correlator $G_{mixed}(\r,\br)=\langle \O^2(x_4)  \O^2(x_1)\O(x_2)\O(x_3)\rangle$ for $\r>1$ and $1>\br>0$, where operators are ordered as written. First, consider the contribution of the first term of equation (\ref{tyrion}) to $G_{mixed}(\r,\br)$ which corresponds to the disconnected Witten diagram. Clearly, this contribution is just the analytic continuation of the  Euclidean four-point function (\ref{mixed_G}) of the 5d generalized free theory with $\Delta_\O=5$ and $c_\O=\frac{160}{\pi^3}$. On the other hand, deformations of the generalized free theory  four-point function come from the second and third term of (\ref{tyrion}) which lead to connected Witten diagrams.\footnote{Note that there are partially disconnected Witten diagrams that contribute to the mixed correlator as well. Let us ignore these diagrams for now. We will  later argue that partially disconnected Witten diagrams cannot contribute to $\gamma_3^{\text{Regge}}$. However, these diagrams can contribute to $\gamma_3^{\text{other}}$ but in a very specific way.} The part of the on-shell action (\ref{tyrion}) that contributes to connected Witten diagrams can be simplified to 
\begin{align}\label{six}
S^{(\phi^6)}_{on-shell}&=-\int d^{5}x\ d\z \sqrt{g_{EAdS}(\z)} \mathcal{L}_{int}^0(z,x)\\
+\frac{1}{2}\int &d^{5}xd\z  \sqrt{g_{EAdS}(\z)}\int d^{5}x' d\z' \sqrt{g_{EAdS}(\z')} G(\z,x;\z',x')\frac{\delta \mathcal{L}_{int}^1}{\delta \phi}(\z',x')\frac{\delta \mathcal{L}_{int}^1}{\delta \phi}(\z,x)\nonumber
\end{align}
with
\begin{align}
&\mathcal{L}_{int}^0=\frac{\hat{b}}{128f^{10}}\left(5 \phi^4 \Box^2 \phi^2+2 \phi^3 \Box^2 \phi^3\right)+\frac{\Delta a}{240f^{12}}\left( 45 \phi^4 \Box^3 \phi^2+20 \phi^3 \Box^3 \phi^3\right)\ ,\\
&\frac{\delta \mathcal{L}_{int}^1}{\delta \phi}=\frac{\hat{b}}{2f^4}\left( \Box^2 \phi^2 \right)+\frac{\hat{b}}{4f^6}\left(   \Box^2 \phi^3+\phi\Box^2 \phi^2 \right) +\frac{\hat{b}}{32f^8}\left(5  \Box^2 \phi^4+6\phi^2 \Box^2 \phi^2+4\phi \Box^2 \phi^3\right)\ , \nonumber
\end{align}
where, we dropped all terms that contribute only at an order higher than $\frac{1}{\Delta_f^{12}}$. At the order $\frac{1}{\Delta_f^{12}}$,
 both exchanged and contact Witten diagrams contribute to the connected deformation of the mixed correlator  $\delta G_{mixed}(\r,\br)$ which will be discussed at length below.

 \subsubsection{Exchanged diagrams}
At first sight, it may appear that there are many different terms that contribute to the exchanged Witten diagram. However, most of the terms can be converted to contact diagrams by integration by parts. In fact,  there is a single term of the on-shell action (\ref{six}) that truly corresponds to an exchanged Witten diagram
\be\label{action4}
\frac{\hat{b}^2}{32f^{12}}\int d^{5}xd\z  \sqrt{g_{EAdS}(\z)}\int d^{5}x' d\z' \sqrt{g_{EAdS}(\z')}\ \phi\(\Box^2 \phi^2\) G(\z,x;\z',x')\ \phi'{\Box'}^2 {\phi'}^2 ,
\ee
where, we are using the notation $\phi\equiv \phi(\z,x)$ and $\phi'\equiv \phi(\z',x')$. The leading contribution of this term to the mixed correlator $G_{mixed}(\r,\br)=\langle \O^2(x_4)  \O^2(x_1)\O(x_2)\O(x_3)\rangle$ in the Regge limit (\ref{regge2}) is  given by the Witten diagram 
\be\label{diagram1}
\includegraphics[scale=0.5]{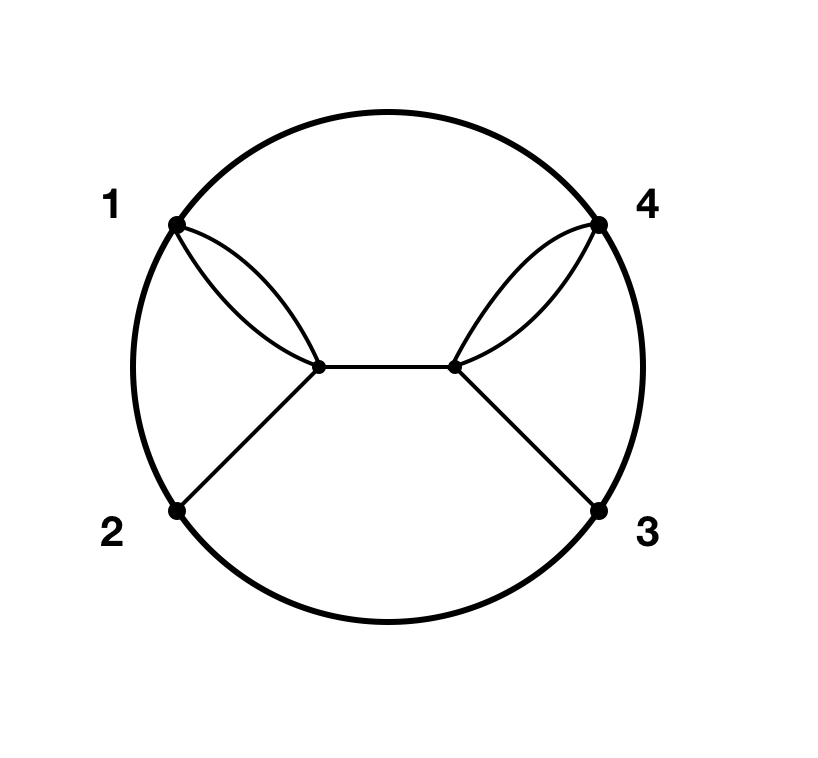}\ .
\ee
The Regge behavior of this diagram can be analyzed by using the identity (\ref{identity1}) yielding 
\begin{align}
\Box^2 \(\tilde{K}_{5}(\z,x;x_i)\tilde{K}_{5}(\z,x;x_j)\)=&4\(5\)^4\tilde{K}_{5}(\z,x;x_i)\tilde{K}_{5}(\z,x;x_j)\nonumber\\
&-536(5)^2  x_{ij}^2\tilde{K}_{6}(\z,x;x_i)\tilde{K}_{6}(\z,x;x_j)\nonumber\\
&~~~+(4)^2(5)^2(6)^2 x_{ij}^4\tilde{K}_{7}(\z,x;x_i)\tilde{K}_{7}(\z,x;x_j)\ .
\end{align}
This identity greatly simplifies the exchanged diagram (\ref{diagram1}) by transforming it into a finite sum over four-point scalar-exchanged Witten diagrams $W_{\Delta_1\Delta_2 \Delta_3 \Delta_4}$ (\ref{defineW}). In particular, the exchanged diagram (\ref{diagram1}) for the kinematics (\ref{points}) only contains terms $(\r\br)^{m_1}W_{10+m_1,5+m_1,5+m_2,10+m_2}(\r,\br)$ with $m_1,m_2=0,1,2$. In general $W$-functions are complicated objects. Since, however, any four-point scalar-exchanged Witten diagram can be decomposed into (infinite) sums over only {\it scalar}-exchanged conformal blocks \cite{Hijano:2015zsa}, the Regge behavior of $W$-functions can be obtained even without trying. In the Regge limit (\ref{regge2}) followed by $\r\br\rightarrow 0$, using (\ref{rgblock}) for $\ell=0$ along with (\ref{G2}) we conclude that at the leading order in $\rho$
\be
(\r\br)^{m_1}W_{10+m_1,5+m_1,5+m_2,10+m_2}(\r,\br)\sim i \frac{(\r\br)^{m}}{\r}
\ee
for some real $m$. Hence, the contribution of the on-shell action (\ref{action4}) to $\delta G_{mixed}(\r,\br)$ can never grow as $\r^2$ in the Regge limit implying that exchanged diagrams cannot contribute to $\gamma_3^{\text{Regge}}$.

The on-shell action (\ref{action4})  still can contribute to $\gamma_3^{\text{other}}$ through scalar exchanged Witten diagrams in other channels. Unfortunately, this contribution is difficult to calculate analytically. However, from (\ref{action4}) it is clear that any such contribution can be written as 
\be
\gamma_3^{\text{exchange}}=\alpha_{\text{exchange}} \gamma_2^2
\ee
for any 6d RG flow where $\alpha_{\text{exchange}} $ is a model independent numerical factor. Moreover, in the standard scenario in which  RG flows  are triggered by adding  relevant (or marginally relevant) operators that break  conformal symmetry explicitly, we can ignore $\gamma_3^{\text{exchange}}$ since $\hat{b}\ll \Delta a$. 

\subsubsection{Contact diagrams}
This brings us to the  contact diagram
\be\label{diagram2}
\includegraphics[scale=0.5]{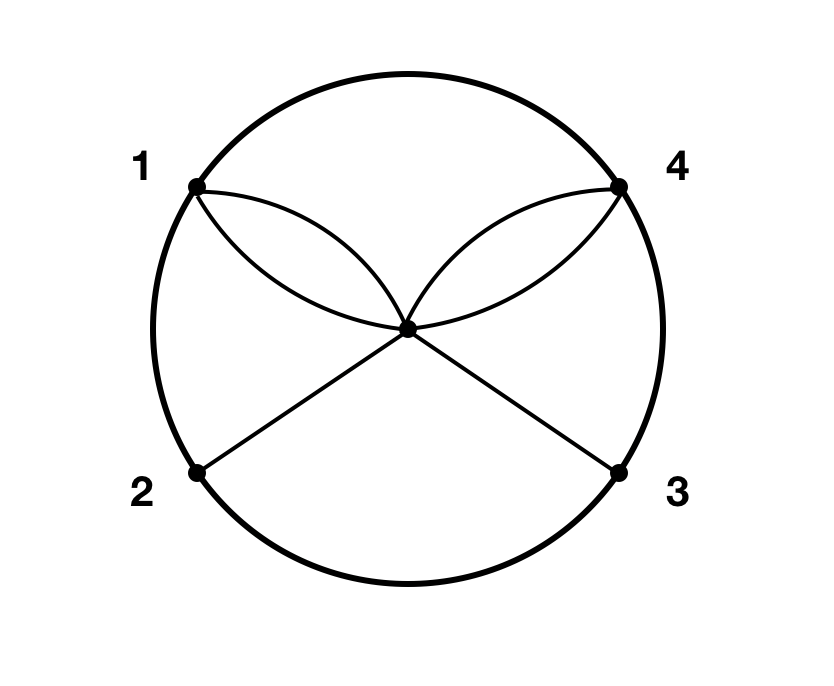}\ .
\ee
There are various vertices that contribute to this contact diagram. To see that we organize the part of the on-shell action (\ref{six}) that contributes to the contact diagram in the following way 
\be
S^{(\phi^6)}_{contact}=S^{(\phi^6)}_{(4)}+S^{(\phi^6)}_{(6)}
\ee
where at the four-derivative order, after using the equation of motion, we have 
\be
S^{(\phi^6)}_{(4)}=-\frac{3\hat{b}}{8f^{10}}\int_{AdS_6} \phi^2 (\Box \phi^2)^2 \ .
\ee
On the other hand, there are two separate contact interactions at the six derivative level
\be\label{action6}
S^{(\phi^6)}_{(6)}=-\frac{8\Delta a-5\hat{b}^2}{96f^{12}} \int_{AdS_6}\ \phi^3 \Box^3 \phi^3-\frac{3\Delta a-2\hat{b}^2}{16f^{12}} \int_{AdS_6} \phi^4 \Box^3 \phi^2\ ,
\ee
where, we have again used the equation of motion to simplify the on-shell action. Notice that the second term of (\ref{six}) has contributed to (\ref{action6}) at the order $\frac{\hat{b}^2}{f^{12}}$. There are terms in $\frac{\delta \mathcal{L}_{int}^1}{\delta \phi}$ that are total-derivative. Contributions of these terms in  (\ref{six}), after integration by parts, reduce to contact interactions. 

The observant reader may have noticed that combinations $(3\Delta a-2\hat{b}^2)$ and $(8\Delta a-5\hat{b}^2)$ also appear in the four-point, five-point, and six-point on-shell dilaton matrix elements at the order $p^6$ (see \cite{Elvang:2012st}). Of course, this is not surprising since there is a one-to-one correspondence between flat space dilaton S-matrix elements and crossing-symmetric Mellin space CFT correlators of deformed generalized free theory.

Let us now determine if the contact diagram (\ref{diagram2}) contributes to $\gamma_3$. First, consider the four-derivative action $S^{(\phi^6)}_{(4)}$. As the discussion of the previous section leads us to expect, four-derivative interactions do not contribute to $\gamma_3$. This is rather easy to show. The on-shell action $S^{(\phi^6)}_{(4)}$ can be expanded by using the bulk-to-boundary propagator 
\begin{align}\label{ma4}
S^{(\phi^6)}_{(4)}=-\frac{3 \hat{b}}{2f^{10}}\  \left( \frac{32}{\pi ^3}\right)^6 5^4\int_{\Phi^6}\int_{AdS_6}&\tilde{K}_{5}(y_6)\tilde{K}_{5}(y_5)\left(\tilde{K}_{5}(y_3)\tilde{K}_{5}(y_4)-2 y_{34}^2 \tilde{K}_{6}(y_3)\tilde{K}_{6}(y_4) \right)\nonumber\\
&\times \left(\tilde{K}_{5}(y_1)\tilde{K}_{5}(y_2)-2 y_{12}^2 \tilde{K}_{6}(y_1)\tilde{K}_{6}(y_2) \right)\ ,
\end{align}
where we are using the simplified notation (\ref{notation}). The rest of the argument is exactly the same as section \ref{spin2}. The  contribution of (\ref{ma4}) to $G_{mixed}(\r,\br)$  in the Regge limit (\ref{regge2}) followed by the limit $\r\br\rightarrow 0$ is 
\be
\delta G_{mixed}(\r,\br)\sim i \frac{ \hat{b}}{\Delta_f^{10}} \r \log\(\r\br\)
\ee
implying that the four-derivative six-point interactions do not contribute to $\gamma_3$ as expected. However, they do contribute to anomalous dimensions of spin-2 triple-trace operators $[\O^3]_{n,2}$. 

Notice that the leading Regge contribution always comes from terms in the on-shell action with the highest number of $y_{ij}^2$ factors. This follows from the fact that all $D$-functions have the same Regge behavior $D_{\Delta_1\Delta_2 \Delta_3 \Delta_4}(\r,\br)\sim\frac{1}{\r}$. So, we  use this fact to simplify the six-derivative six-point interaction $S^{(\phi^6)}_{(6)}$
\begin{align}\label{ma5}
S^{(\phi^6)}_{(6)}&=\frac{8\Delta a-5\hat{b}^2}{f^{12}} \left( \frac{32}{\pi ^3}\right)^6 (2)^4 (3)^3 (5)^3  \int_{\Phi^6}\int_{AdS_6}   y_{13}^2 y_{12}^2\tilde{K}_{5}(y_6)\tilde{K}_{5}(y_5)\tilde{K}_{5}(y_4)\tilde{K}_{7}(y_2)\nonumber\\
&\qquad \qquad \qquad ~~~~~~~~~~~~~~~~~~~~~~ \times \(7 y_{12}^2 \tilde{K}_{8}(y_1)\tilde{K}_{6}(y_3)+2 y_{23}^2\tilde{K}_{7}(y_1)\tilde{K}_{7}(y_3) \)\nonumber\\
&+\frac{14a-9\hat{b}^2}{f^{12}} \left( \frac{32}{\pi ^3}\right)^6 2(5)^2(6)^2 (7)^2 \int_{\Phi^6}\int_{AdS_6}  y_{12}^6 \tilde{K}_{5}(y_6)\tilde{K}_{5}(y_5)\tilde{K}_{5}(y_4)\tilde{K}_{5}(y_3)\tilde{K}_{8}(y_2)\tilde{K}_{8}(y_1)\nonumber\\
&+\cdots \ ,
\end{align}
where, dots represent terms that can not contribute at the order $\r^2$. Let us now compute the  leading Regge contribution from the on-shell action (\ref{ma5}) to the Lorentzian correlator $G_{mixed}(\r,\br)=\langle \O^2(x_4)  \O^2(x_1)\O(x_2)\O(x_3)\rangle$ in  the kinematics (\ref{points}), where operators are ordered as written. It takes long but straightforward algebra to confirm that the six-derivative six-point interactions do contribute at the order $\r^2$. In particular, we obtain\footnote{We have used the fact that $D$-functions in the Regge limit obey  the property $D_{\Delta_1 \Delta_2\Delta_3 \Delta_4}(\r,\br)=D_{\Delta_4 \Delta_3\Delta_2 \Delta_1}(\r,\br)$. Furthermore, we suppressed the dependence of $D$-functions on $\r$ and $\br$ to lighten the notation.} 
\begin{align}
\delta G_{mixed}(\r,\br)&= -\frac{\r^3}{ \Delta_f^{12}}\left( \frac{32}{\pi ^3}\right)^6 7(3)^3 (5)^2  (2)^5  \left[28 \left(3 \Delta a-2 b^2\right)\left(4 D_{10,8,5,13}-D_{10,8,8,10}-4 D_{13,5,5,13}\right) \right.\nonumber\\
&+\left(8 \Delta a-5 b^2\right)\left(86 D_{10,8,5,13}+30 D_{10,8,7,11}-14 D_{10,8,8,10} \right.\nonumber\\
&\left.\left.+60 D_{12,6,5,13}-30 D_{12,6,8,10}-116 D_{13,5,5,13}-60 D_{13,5,7,11}\right) \right]
\end{align}
where the $D$-function is defined in (\ref{defineD}). The above expression is not very transparent. So, we simplify it further by taking the limit: $\r\br\rightarrow 0$. In this limit, the we can use the analytic result (\ref{finalD}) to obtain
\be\label{corr6}
\delta G_{mixed}(\r,\br)\approx - i  \frac{4\Delta a-3\hat{b}^2}{ \Delta_f^{12}}\(\frac{32}{\pi ^3}\right)^6\frac{(5)^4 (3)^2  \pi ^{\frac{7}{2}}  \Gamma \left(\frac{31}{2}\) \Gamma (17)}{2^{35} \Gamma (13) \Gamma (12)} \r^2 \log\(\r\br\)\ .
\ee
Clearly, this contribution can only come from the anomalous dimension of the operator $[\O^3]_{0,\ell=3}$. Moreover, comparing the above result with (\ref{formula2}), we can easily identify $\gamma_3$
\be\label{gamma3}
\gamma_3^{\text{contact}}=\frac{\alpha_{\text{contact}}}{\Delta_f^{12}}\(\frac{51}{8 \pi ^3}\)^2 \(- \frac{4}{3}\Delta a+\hat{b}^2\)\ , \qquad \alpha_{\text{contact}}=\(\frac{3}{4}\)^2\frac{\Gamma \left(\frac{31}{2}\right) \Gamma \left(\frac{15}{2}\right)^2}{\Gamma \left(\frac{21}{2}\right) \Gamma \left(\frac{19}{2}\right)^2}
\ee
where, we have used equation (\ref{lambda}) and the generalized free field value of $\tilde{c}(0,3)^2=\frac{14000}{323}c_0^3$. Thus we conclude that $\Delta a$ contributes to the anomalous dimension of the spin-3 triple-trace operator $[\O^3]_{0,3}$ of the dual CFT$_5$ at the leading order. 

As we remarked earlier, anomalous dimensions of triple-trace operators are complicated objects. For simplicity, we can always focus on the part of $\gamma_3$ that controls the Regge growth (\ref{growth}). This part of $\gamma_3$ is completely fixed by the above contact diagram implying
\begin{bBox}
\be\label{gammaregge}
\gamma_3^{\text{Regge}}=\frac{\alpha_{\text{contact}}}{\Delta_f^{12}}\(\frac{51}{8 \pi ^3}\)^2 \(- \frac{4}{3}\Delta a+\hat{b}^2\)\ .
\ee
\end{bBox}
Note that $\gamma_3$ and $\gamma_3^{\text{Regge}}$ both  receive contributions also from $\hat{b}^2$ at the same order. Hence, $\gamma_3$ and $\gamma_3^{\text{Regge}}$ of the dual CFT$_5$ depend on the details of the 6d RG flow. However, there is a specific combination of $\gamma_3$ (or $\gamma_3^{\text{Regge}}$) and $\gamma_2^2$ which depends only on the UV and IR fixed points. Before we discuss that there are a few loopholes in our argument that we must address. 

The correlator (\ref{corr6}), strictly speaking, implies (\ref{gamma3}) if and only if there is a unique lowest twist spin-3 triple-trace operator.  In general, as noted earlier, a multi-trace operator  $[\O^N]_{n,\ell}$ can be degenerate. However, one might expect that $[\O^N]_{n,\ell}$ is non-degenerate for sufficiently small $n$ and $\ell$. This is certainly true for $[\O^3]_{0,\ell}$ with $\ell\le 3$ which we have established by explicitly constructing them for the generalized free theory in appendix \ref{tripletrace}.

\subsubsection{Partially disconnected diagrams}
Consider the partially disconnected Witten diagram 
\be\label{diagram3}
\includegraphics[scale=0.5]{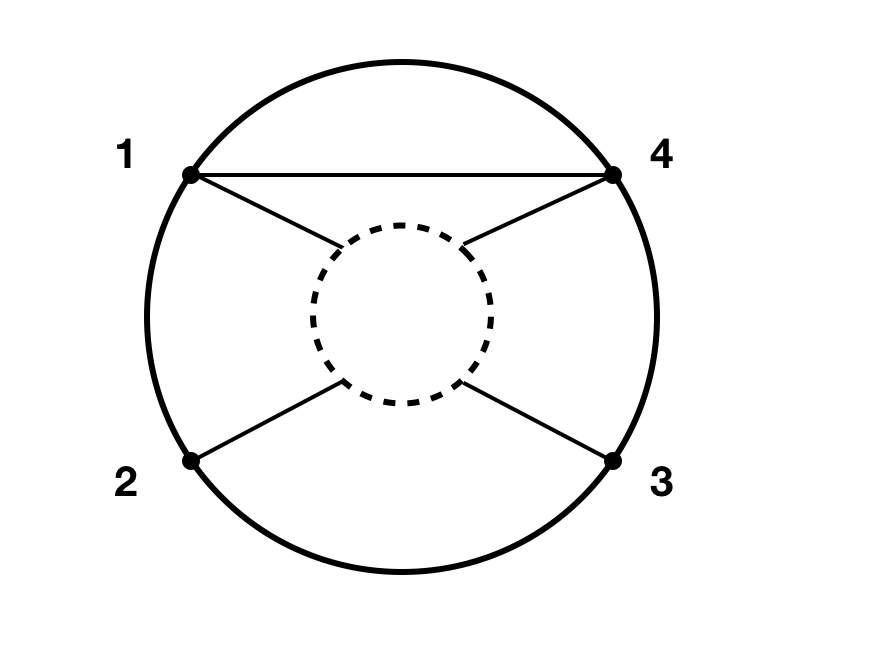}
\ee
which includes loop effects as well. Clearly, this is the only partially disconnected Witten diagram that can potentially contribute to $\gamma_3$.  One might expect that the one-to-one correspondence between bulk interactions and anomalous dimensions of multi-trace operators implies that bulk interactions of the diagram  (\ref{diagram3}) can never contribute to $\gamma_3$. This is certainly true at tree level, however, this argument is not valid once we include loops or partially disconnected Witten diagrams. To be specific, contribution of the above Witten diagram can be written as
\be
\delta G_{mixed}(\r,\br)= \frac{2c_\O}{\(1-\r\)^5(1-\br)^5}\delta G_4(\r,\br)\ .
\ee
Conformal block expansions of both $G_{mixed}(\r,\br)$ and $G_4(\r,\br)$ now imply that $\gamma^{(3)}_{n,\ell}$ does receive contributions from $\gamma^{(2)}_{n,\ell}$. In particular, at the leading order we obtain
\be
\gamma_3^{\text{disc}}=\frac{51}{22}\gamma_2\ .
\ee
Hence, a part of $\gamma_3$ comes entirely from the anomalous dimension of the double-trace operator $[\O^2]_{0,2}$. Alternatively, the binding energy of a spin-$\ell$ three-particle state in AdS always receives a contribution from purely two-particle bound states with spin $\le \ell$. So, we define a subtracted 
\be
\delta \gamma_3\equiv \gamma_3-\frac{51}{22}\gamma_2
\ee
which corresponds to  the true  three-particle  binding energy. At the next order in perturbation theory, both $\gamma_3$ and $\gamma_2$ are UV divergent because of loop diagrams. However, the combination $\delta \gamma_3$ is finite and hence scheme-independent. In addition, at the subleading order $\gamma_3^{\text{disc}}$ receives a correction which we will denote as $\alpha_{\text{disc}}\gamma_2^2$, where $\alpha_{\text{disc}}$ is a finite model independent numerical coefficient.

In the beginning of this section, we claimed that partially disconnected Witten diagrams, even when we consider loops, cannot contribute to $\gamma_3^{\text{Regge}}$. It is rather easy to establish that claim by utilizing (\ref{loop}). In particular, at the order $\frac{1}{\Delta_f^{12}}$ equation (\ref{loop}) dictates  that the contribution of (\ref{diagram3}) to $\delta G_{mixed}(\r,\br)$ cannot grow in the Regge limit faster than
\be
\delta G_{mixed}(\r,\br)\sim  \frac{\r^{-5}}{\(1-\frac{1}{\r}\)^5(1-\br)^5}\delta G^{1-loop}_4(\r,\br)\approx i \frac{\r^{-2}}{\Delta_f^{12}}
\ee
implying partially disconnected Witten diagrams cannot contribute to $\gamma_3^{\text{Regge}}$.

~\\

Let us now put everything together. The total $\gamma_3$, after adding all the contributions, is given by\footnote{The actual value of $\Delta_f$ is not important. So, we have defined $\tilde{\Delta}_f$ such that it absorbs all the prefactors in (\ref{gamma3}): 
\be \frac{3}{4}\alpha^2 \(\frac{51}{8\pi^3}\)^2\tilde{\Delta}_f^{12}=\Delta_f^{12}\ .
\ee}  
\begin{bBox}
\be\label{total}
\delta \gamma_3= \gamma_3-\frac{51}{22}\gamma_2=-\frac{\Delta a}{\tilde{\Delta}_f^{12}}+\alpha \gamma_2^2\ , 
\ee
\end{bBox}
where, $\alpha=\alpha_{\text{exchange}} +\alpha_{\text{contact}}+\alpha_{\text{disc}}$ is a universal numerical coefficient which does not depend on the details of the flow. The exact value of $\alpha$ can be computed numerically, however, we will not attempt it in this paper. This concludes our discussion of triple-trace operators.

\addtocontents{toc}{\protect\setcounter{tocdepth}{1}}
\section{Conclusions \& Discussion}\label{section_conclusions}
\addtocontents{toc}{\protect\setcounter{tocdepth}{-1}}

RG flows connecting two conformal fixed points can be described by the dilaton effective action of  broken conformal symmetry.  In this paper, we have analyzed the dilaton effective action in AdS by studying the dual CFT. The dual CFT, for any finite but large $R_{\text{AdS}}$, must be well behaved in the usual sense. This is particularly useful in even spacetime dimensions where $\Delta a$ can be related to anomalous dimensions of lowest twist multi-trace operators. For example, the proof of the $a$-theorem in 4d by Komargodski and Schwimmer can be reinterpreted as a CFT$_3$ statement $\Delta a= - \tilde{\Delta}_f^4 \gamma_2\ge 0$. RG flows in 6d are more subtle. However, we can still map the 6d $a$-theorem for RG flows from spontaneously broken conformal symmetry into a statement about anomalous dimensions in the dual CFT$_5$ by utilizing relations (\ref{gamma2}), (\ref{gamma3}) and (\ref{total})
\be\label{final1}
\frac{\Delta a}{{\tilde{\Delta}_f}^{12}}=-\delta \gamma_3+\alpha \gamma_2^2=  -\gamma_3^{\text{Regge}}+\alpha_{\text{contact}} \gamma_2^2\ , \qquad  \alpha_{\text{contact}}=\(\frac{3}{4}\)^2\frac{\Gamma \left(\frac{31}{2}\right) \Gamma \left(\frac{15}{2}\right)^2}{\Gamma \left(\frac{21}{2}\right) \Gamma \left(\frac{19}{2}\right)^2}
\ee
where, $\alpha$ and $ \alpha_{\text{contact}}$ are model independent numerical constants that are completely fixed by conformal symmetry of the dual description. This is our main result. The fact that all interactions of the dilaton effective theory are non renormalizable implies that a smooth flat space limit $R_{\text{AdS}}\rightarrow \infty$ exists for the AdS dilaton action. This guarantees that the positivity of the right hand side of (\ref{final1}) for the deformed generalized free theory in 5d is sufficient to establish the 6d $a$-theorem.

It is more natural to consider  the case where the $\CFTUV$ is deformed by a relevant or marginally relevant operator. This breaks conformal symmetry explicitly which triggers an RG flow to some $\CFTIR$. Every RG flow with explicit symmetry breaking can be thought of as a special case of  RG flows with spontaneous symmetry breaking with $\hat{b}\ll 1$. This immediately implies 
\be\label{final3}
\frac{\Delta a}{{\tilde{\Delta}_f}^{12}}=-\delta \gamma_3=  -\gamma_3^{\text{Regge}}\ .
\ee
for explicitly broken conformal symmetry in 6d.

\subsubsection*{Anomalous dimensions of $\O$ and $\O^2$}
In the derivation of (\ref{final1}), there was an implicit assumption that both $\O$ and $\O^2$ do not acquire any anomalous dimension as a result of the RG flow. A related but slightly stronger statement is that the flow does not generate a potential for the dilaton field $\phi$. Even if we start without a potential for the dilaton, generally the flow can generate   a cosmological constant term in (\ref{dilaton}) particularly when conformal symmetry is broken explicitly. However, this IR cosmological constant term can be removed by adding an appropriate bare cosmological term in (\ref{dilaton}). This implies that anomalous dimensions of $\O$ and $\O^2$ can always be tuned to zero. In fact, any CFT based analysis of the relation (\ref{final1}) must set $\Delta_\O=5$ and $\Delta_{\O^2}=10$.

\subsubsection*{A CFT sum rule}
The relation (\ref{gamma3}) can be equivalently written as a CFT dispersion sum rule
\begin{align}\nonumber
\frac{\Delta a}{{\tilde{\Delta}_f}^{12}}=\lim_{\eta\rightarrow 0}\lim_{x\rightarrow 0}\left[\frac{1}{\log \eta} \int^x_{-x} d\sigma\ \sigma\ \mbox{Re}\ \delta G_{mixed}\(\eta, \sigma\)+\frac{\beta^2}{\(\log \eta\)^2}\( \int^x_{-x} d\sigma\ \mbox{Re}\ \delta G_4\(\eta, \sigma\)\)^2\right]
\end{align}
where $G(\eta,\sigma)\equiv G(\r=\frac{1}{\sigma},\br=\eta \sigma)$ and $\beta^2=\frac{25346}{503965}$ is a universal numerical factor. Moreover, we have also absorbed a positive numerical coefficient in the definition of $\tilde{\Delta}_f$. The above sum rule follows from analyticity of Lorentzian CFT correlators which implies that both  $\delta G_{mixed}$ and $\delta G_4$, as functions of complex $\sigma$, are analytic in the lower half $\sigma$-plane near $\sigma\sim 0$ \cite{Hartman:2016lgu}. However, this CFT sum rule is not manifestly positive definite and hence it does not immediately lead to a proof of the 6d $a$-theorem.

\subsubsection*{Massive scalar field theory}
A simple example of explicitly broken conformal symmetry is given by the free massive scalar field theory in 6d. In this case, the $\CFTUV$ is a massless scalar field theory and hence $a_{UV}$ is known exactly \cite{Elvang:2012st,Bastianelli:2000hi}
\be
a_{UV}=\frac{1}{(4\pi)^3 9072}\ .
\ee
Now conformal symmetry can be broken explicitly by introducing a mass term for the scalar field. Clearly, the scalar field decouples in the deep  IR and the theory flows to nothing. Therefore, in this case $\gamma_3$ is given by (\ref{final3}) with $\Delta a=a_{UV}$. The dilaton effective theory associated with this RG flow was analyzed in detail in \cite{Elvang:2012st} which led to an exact result for $\hat{b}$
\be
0<\hat{b}=\frac{m^2}{(4\pi)^3 360 f^2}\ll 1\ ,
\ee
where, $m$ is the mass of the scalar field.

\subsubsection*{Supersymmetry and the $a$-theorem}
All known examples of interacting CFTs in 6d have one thing in common -- they are supersymmetric. So, naturally RG flows connecting two SCFTs are of significant importance in 6d. For conformal UV fixed points that are supersymmetric, superconformal representation theory dictates  that conformal symmetry of the SCFT$_{\text{UV}}$ cannot be broken explicitly in 6d while preserving supersymmetry \cite{Cordova:2016xhm,Cordova:2016emh,Louis:2015mka}. Spontaneously broken conformal symmetry can  set off flows that preserve supersymmetry, however, such flows are highly constrained. Constraints imposed by supersymmetry were nicely exploited to establish the 6d $a$-theorem for all flows that preserve $(2,0)$ supersymmetry in \cite{Cordova:2015vwa} which was later extended to RG flows of $(1,0)$ SCFTs onto the tensor branch in \cite{Cordova:2015fha}. The $a$-theorem for this class of theories follows from the fact that supersymmetry relates $\Delta a$ to $\hat{b}^2$. In particular, for these supersymmetric flows  \cite{Cordova:2015vwa, Cordova:2015fha}\footnote{Note that our convention for $a$ and $\hat{b}$ is different from the convention used in \cite{Cordova:2015fha}. }
\be
\Delta a =\frac{2\hat{b}^2}{3} \ge 0\ .
\ee
It is a straightforward exercise to rewrite this result as statements about anomalous dimensions $\gamma_2$ and $\gamma_3^{\text{Regge}}$ of the dual CFT$_5$
\be\label{didi1}
\frac{\Delta a}{\tilde{\Delta}_f^{12}}= \frac{8}{9} \alpha_{\text{contact}} \gamma_2^2\ge 0\ , \qquad \gamma_3^{\text{Regge}}=\frac{1}{9} \alpha_{\text{contact}} \gamma_2^2\ .
\ee
Interestingly, the results (\ref{final3}) and (\ref{didi1}) imply that a general Nachtmann-like CFT theorem that strictly rules out either sign of $\gamma_3^{\text{Regge}}$ cannot exist. 

In a $(1,0)$ SCFT, there are two types of deformations that preserve supersymmetry.  These are tensor branch flows and Higgs branch flows. The classification of 6d SCFTs  has provided strong evidence for the $a$-theorem even for Higgs branch flows of $(1,0)$ SCFTs \cite{Heckman:2015axa,Heckman:2016ssk,Heckman:2018jxk}. However, it is still an open problem to establish the $a$-theorem for all RG flows of $(1,0)$ SCFTs onto the Higgs branch.

\section*{Acknowledgements}
It is our pleasure to thank Jared Kaplan for several helpful discussions as well as comments on a draft. We would also like to thank Nima Afkhami-Jeddi, Ibou Bah, Federico Bonetti, David Simmons-Duffin, Thomas Dumitrescu, Tom Hartman, and Jonathan Heckman for helpful discussions.  We were supported in part by the Simons Collaboration Grant on the Non-Perturbative Bootstrap.

\appendix
\addtocontents{toc}{\protect\setcounter{tocdepth}{-1}}

\section{Conformal trace anomaly in 6d}\label{app_anomaly}
The conformal trace anomaly in $d=6$ can be written as \cite{Bonora:1985cq, Deser:1993yx, Bastianelli:2000hi,Boulanger:2007ab}
 \be\label{anomaly}
 \langle T^\mu_\mu\rangle = a E_6+ \sum_{i=1}^3 c^{(i)} I_i\ 
 \ee
 up to total derivative terms which can be removed by adding finite and covariant counter-terms in the effective action. In equation (\ref{anomaly}) $E_6$ is the 6d Euler density\footnote{We are using the convension of  \cite{Elvang:2012st}.}
 \be
 E_6= \frac{1}{8}\delta^{\mu_1 \mu_2 \mu_3 \mu_4 \mu_5 \mu_6}_{\nu_1 \nu_2 \nu_3 \nu_4 \nu_5 \nu_6}R^{\nu_1\nu_2}_{~~~~\mu_1 \mu_2}R^{\nu_3\nu_4}_{~~~~\mu_3 \mu_4}R^{\nu_5\nu_6}_{~~~~\mu_5 \mu_6}
 \ee
 and $a$ is the corresponding Euler central charge. On the other hand, central charges $\{c^{(i)}\}$ are associated with conformal invariants:
 \begin{align}
 I_1=& W_{\gamma \alpha \beta \delta}W^{\alpha \mu \nu \beta}W_{\mu~~\nu}^{~\gamma \delta}\ , \notag\\
 I_2=& W_{\alpha \beta}^{~~\gamma \delta}W_{\gamma \delta}^{~~\mu \nu}W_{\mu \nu}^{~~\alpha \beta}\ , \notag\\
 I_3=& W_{\alpha \gamma \delta \mu}\left(\nabla^2 \delta^\alpha_\beta+4 R^\alpha_\beta-\frac{6}{5}R \delta^\alpha_\beta \right)W^{\beta \gamma \delta \mu}\ ,
 \end{align}
where, $W$ is the Weyl tensor.

\section{Analytic continuation of hypergeometric functions}
Hypergeometric function ${}_2F_1 \left(a,b,c,z\right)$ has a branch cut along $z \in (1,\infty)$. If we start with $0<z<1$ and rotate $z$ around 1: $(1-z)\rightarrow (1-z)e^{-2\pi i}$, we obtain the following identity 
\begin{align}\label{identity}
{}_2F_1 &\left(a,b,c,z\right)_{(1-z)\rightarrow (1-z)e^{-2\pi i}}={}_2F_1\left(a,b,c,z\right)\\
&+\frac{2\pi i \Gamma(c) e^{-\pi i(c-b-a)}}{\Gamma(a)\Gamma(b)\Gamma(c-a-b+1)}\frac{(1-z)^{c-b-a}}{z^{c-1}}\ _2 F_1 (1-b,1-a,c-a-b+1,1-z)\ .\nonumber
\end{align}

\section{Feynman rules for the dilaton in AdS}\label{holography}
In this appendix, we lists all the Feynman rules that we will use to evaluate various Witten diagrams. We will use the following convention for points in AdS$_6$: $(\z,x)$,  where $\mathfrak{ z}$ is the bulk direction and $x\in \mathbb{R}^5$. For convenience we will work in the Euclidean signature with the metric
\be\label{metric}
ds^2=\frac{d\z^2+dx^2}{\z^2}\ .
\ee
\subsubsection*{Bulk-to-boundary propagator}
The dilaton bulk-to-boundary propagator between a bulk point $(\z, x)$ and a boundary point $x'$ in Euclidean AdS$_6$ is given by
\be
K(\mathfrak{z},x;x')=\left( \frac{32}{\pi ^3}\right)\frac{\z^5}{(\z^2+|x-x'|^2)^5}\ .
\ee
For notational convenience, we also define a reduced bulk-to-boundary propagator
\be\label{reduced}
\tilde{K}_{\Delta}(x')\equiv \tilde{K}_{\Delta}(\mathfrak{z},x;x')=\frac{\z^\Delta}{(\z^2+|x-x'|^2)^\Delta}\ .
\ee
\subsubsection*{Bulk-to-bulk propagator}
The dilaton bulk-to-bulk propagator is the solution of the differential equation
\be 
\Box(\z,x)G(\z,x;\z',x')=\frac{1}{\sqrt{g_{EAdS}(\z)}}\delta(\z-\z')\delta^d(x-x')\ .
\ee
The propagator can  be explicitly written as
\be
G(\z,x;\z',x')=-\(\frac{\xi ^{5 } }{5 \pi^3}\)\, _2F_1\left(\frac{5 }{2},3;\frac{7}{2};\xi ^2\right)
\ee
where,
\be
\xi=\frac{2\z  \z'}{\z^2+{\z'}^2+(x-x')^2}\ .
\ee
\subsubsection*{Contact Witten diagram: $D$-functions}
We define the $D$-function in AdS$_{d+1}$ in the traditional way 
\be\label{defineD}
D_{\Delta_1\Delta_2 \Delta_3 \Delta_4}(x_1,x_2,x_3,x_4)=\int d^{d}x\ d\z \sqrt{g_{AdS}(\z)} \prod_{i=1}^4 \tilde{K}_{\Delta_i}(\z,x;x_i)
\ee
where, $\tilde{K}$ is the reduced bulk to boundary propagator (\ref{reduced}). $D$-functions appear in four-point contact Witten diagrams.
\subsubsection*{Scalar-exchanged Witten diagram: $W$-functions}
A similar integral appears in scalar-exchanged Witten diagrams which we will denote as the W-function. In AdS$_{d+1}$, the $W$-function is given by
\begin{align}\label{defineW}
W_{\Delta_1\Delta_2 \Delta_3 \Delta_4}(x_1,x_2,x_3,x_4)=&\int d^{d}x\ d\z \sqrt{g_{AdS}(\z)}\int d^{d}x'\ d\z' \sqrt{g_{AdS}(\z)}\nonumber\\
& \times G(\z,x;\z',x')\prod_{i=1}^2 \tilde{K}_{\Delta_i}(\z,x;x_i) \prod_{i=3}^4 \tilde{K}_{\Delta_i}(\z',x';x_i)\ .
\end{align}

\subsubsection*{An useful identity}
The following simple identity will be very useful to us  (see \cite{DHoker:1999kzh})
\begin{align}\label{identity1}
g^{AB}\partial_A \tilde{K}_{\Delta_1}(\z,x;x_1)\partial_B \tilde{K}_{\Delta_2}(\z,x;x_2)=&\Delta_1\Delta_2 \left( \tilde{K}_{\Delta_1}(\z,x;x_1) \tilde{K}_{\Delta_2}(\z,x;x_2) \right.\nonumber\\
&\left.-2x_{12}^2\tilde{K}_{\Delta_1+1}(\z,x;x_1) \tilde{K}_{\Delta_2+1}(\z,x;x_2) \right)\ ,
\end{align}
where, derivatives are taken with respect to bulk coordinates.

\section{Regge limit of the $D$-function in AdS$_{d+1}$}
 The AdS$_{d+1}$ integral in the $D$-function (\ref{defineD}) can be reduced to a single integral of a hypergeometric function \cite{Freedman:1998bj}
\begin{align}\label{calcD}
D_{\Delta_1\Delta_2 \Delta_3 \Delta_4}&(x_1,x_2,x_3,x_4)= \frac{\pi^{d/2}}{2}\frac{\Gamma\(\frac{\sum_i \Delta_i}{2}-\frac{d}{2} \)\Gamma\(\frac{\sum_i \Delta_i}{2}-\Delta_3 \)\Gamma\(\frac{\sum_i \Delta_i}{2}-\Delta_4\)}{\Gamma\(\Delta_1\) \Gamma\(\Delta_2\)\Gamma\(\frac{\sum_i \Delta_i}{2}\)}\frac{(x_{12}^2)^{\Delta_4-\frac{\sum_i \Delta_i}{2}}}{(x_{34}^2)^{\Delta_3}}\nonumber\\
&\times \int_0^\infty d\beta \frac{\(\beta x_{24}^2+x_{14}^2\)^{\Delta_3-\Delta_4}}{\beta^{1-\Delta_2-\Delta_4+\frac{\sum_i \Delta_i}{2}} }\ {}_2F_1 \(-\Delta_4+\frac{\sum_i \Delta_i}{2},\Delta_3,\frac{\sum_i \Delta_i}{2},1-\alpha \)\ ,
\end{align}
where,
\be
\alpha=\frac{\(\beta x_{23}^2+x_{13}^2\)\(\beta x_{24}^2+x_{14}^2\)}{\beta x_{12}^2 x_{34}^2}\ .
\ee
However, we only  need the Regge limit of the $D$-function which can be computed exactly. First note that conformal invariance of the boundary implies that
\be\label{dbar}
D_{\Delta_1\Delta_2 \Delta_3 \Delta_4}(x_1,x_2,x_3,x_4)=\frac{1}{(x_{12}^2)^{\frac{\Delta_1+\Delta_2}{2}}(x_{34}^2)^{\frac{\Delta_3+\Delta_4}{2}}}\left(\frac{x_{14}^2}{x_{24}^2} \right)^{\frac{\Delta_{21}}{2}}\left(\frac{x_{14}^2}{x_{13}^2} \right)^{\frac{\Delta_{34}}{2}} \bar{D}_{\Delta_1\Delta_2 \Delta_3 \Delta_4}(z,\bz)
\ee
where $z,\bz$ are the cross-ratios (\ref{cr}). Since, the $\bar{D}$-function depends only on the cross-ratios, we can determine it starting from any configuration we desire. We will closely follow the configuration used in \cite{Heemskerk:2009pn} to evaluate the Regge limit of the $\bar{D}$-function.

\subsubsection*{Embedding formalism}
It is most convenient to work in the embedding formalism \cite{Penedones:2007ns} which was first proposed by Dirac \cite{Dirac:1936fq}. In this formalism, $(d+1)$-dimensional AdS is embedded in $(d+2)$-dimensional Minkowski space $\mathbb{M}^d\times\mathbb{M}^2$ as follows\footnote{Note that we will set $R_{\text{AdS}}=1$ in this section.}
\be
P^2 =-R_{\text{AdS}}^2\ , \qquad P^0>0\ ,
\ee
where $P=(P^+,P^-,P^a)\in \mathbb{M}^2\times\mathbb{M}^d$ and the $(d+2)$-dimensional metric is $dP^2=-dP^+dP^-+dP^a dP_a$. The conformal boundary of AdS is the space of null rays 
\be
P^2=0\ , \qquad P\sim \lambda P\ .
\ee
Now consider the $D$-function in the embedding space $D_{\Delta_1\Delta_2 \Delta_3 \Delta_4}(P_1,P_2,P_3,P_4)$. The Regge limit can be reached by choosing four points following \cite{Heemskerk:2009pn}:
\begin{align}
&P_1=(1,0,0)\ , \qquad P_3=(\bar{x}^2,1,\bar{x})\ ,\nonumber\\
&P_2=(-1,-x^2,x)\ , \qquad P_4=(0,-1,0)\ ,
\end{align}
with $x^2<0, \bar{x}^2<0$. Note that $P_{ij}=(P_1-P_j)^2=-2P_i\cdot P_j$. The cross-ratios as defined in (\ref{cr}) are
\be
z=\sigma e^s\ , \qquad \bar{z}=\sigma e^{-s}\ 
\ee
where $\sigma=x\bar{x}$ and $\cosh s=-\frac{x.\bar{x}}{x \bar{x}}$. We take $x,\bar{x}\rightarrow 0$ with fixed $s$ to go to the Regge regime. With this choice of kinematics, from (\ref{dbar}) we can write  
\be\label{dbar2}
D_{\Delta_1\Delta_2 \Delta_3 \Delta_4}(P_1,P_2,P_3,P_4)=\frac{e ^{-i\pi\frac{\Delta_{21}+\Delta_{34}}{2}}}{(-x^2)^{\frac{\Delta_1+\Delta_2}{2}}(- \bar{x}^2)^{\frac{\Delta_3+\Delta_4}{2}}} \bar{D}_{\Delta_1\Delta_2 \Delta_3 \Delta_4}(z,\bz)\ ,
\ee
where the branch cut is chosen to be consistent with the analytic continuation \ref{dt1}. It is straightforward to generalize \cite{Heemskerk:2009pn} and write an analytic expression for the Regge limit of 
\be\label{dofp}
D_{\Delta_1\Delta_2 \Delta_3 \Delta_4}(P_1,P_2,P_3,P_4)=\int_{AdS}dX \prod_{i=1}^4 \frac{1}{(-2P_i\cdot X)^{\Delta_i}}\ .
\ee
\subsubsection*{Regge limit}
In the Regge limit, $P_{12}, P_{34}\rightarrow 0$ which simplifies the bulk integral because the dominant contribution to the integral comes from the bulk region which is null separated from both $P_1$ and $P_4$. This can be properly utilized by going to the following AdS coordinates \cite{Heemskerk:2009pn}
\be
X=\(u,v\(1-\frac{u v}{4 \cosh^2r}\),\cosh r \(1-\frac{u v}{2 \cosh^2r}\), \theta_{d-2} \sinh r \)\ ,
\ee
where $\theta_{d-2}\in S^{d-2}$. The hypersurface $u=v=0$ is null separated from  both $P_1$ and $P_4$. In the Regge limit, this hypersurface is also almost null separated from $P_2$ and $P_3$. Hence, the dominant contribution to the integral (\ref{dofp}) comes from the region $|u v|\ll \cosh^2 r$ implying that we can safely approximate $X\approx (u,v,w)$ with $w\in H_{d-1}$. As a consequence, in the Regge limit, we can approximate $D_{\Delta_1\Delta_2 \Delta_3 \Delta_4}(P_1,P_2,P_3,P_4)$ by \cite{Heemskerk:2009pn}
\begin{align}
&i \int\frac{dudv}{2}\int_{H_{d-1}} \frac{dw}{(v+i\epsilon)^{\Delta_1} (-v-2x.w+i \epsilon)^{\Delta_2} (u-2\bar{x}.w+i \epsilon)^{\Delta_3}(-u+i\epsilon)^{\Delta_4}}\\
&=2\pi^2 i \frac{\Gamma(\Delta_1+\Delta_2-1)\Gamma(\Delta_3+\Delta_4-1)}{\prod_i \Gamma(\Delta_i)}\int_{H_{d-1}}\frac{dw}{(-2\bar{x}.w+i \epsilon)^{\Delta_3+\Delta_4-1} (-2x.w+i \epsilon)^{\Delta_1+\Delta_2-1}}\ ,\nonumber
\end{align}
where, we are using the standard $i\epsilon$-prescription to implement $(1-z)\rightarrow (1-z)e^{-2\pi i}$ and the factor of $i$ comes from the Wick rotation of the bulk time coordinate. Harmonic analysis on hyperbolic space\footnote{For a review see \cite{Cornalba:2007fs}.}  enables us to evaluate the above integral yielding
\begin{align}\label{bck1}
D_{\Delta_1\Delta_2 \Delta_3 \Delta_4}(P_1,P_2,P_3,P_4)=\frac{i\pi^d}{2|x|^{\Delta_1+\Delta_2-1}|\bar{x}|^{\Delta_3+\Delta_4-1}\prod_i \Gamma(\Delta_i) }f_{\Delta_1\Delta_2 \Delta_3 \Delta_4}(s)\ ,
\end{align}
where, 
\begin{align}
f_{\Delta_1\Delta_2 \Delta_3 \Delta_4}(s)=&\int_{-\infty}^\infty d\nu \Omega_{i\nu}(s)\Gamma\left(\frac{\Delta_3+\Delta_4-d/2+i\nu}{2} \right) \Gamma\left(\frac{\Delta_3+\Delta_4-d/2-i\nu}{2} \right)\nonumber\\
&\times\Gamma\left(\frac{\Delta_1+\Delta_2-d/2+i\nu}{2} \right)\Gamma\left(\frac{\Delta_1+\Delta_2-d/2-i\nu}{2} \right)\ .
\end{align}
Harmonic functions $\Omega_{i\nu}$ on $H_{d-1}$  are known in any dimension \cite{Costa:2017twz}
\begin{align}
\Omega_{E}\left(s\right)=&-\frac{E \sin(\pi E) \Gamma\left(\frac{d-2}{2}+E \right)\Gamma\left(\frac{d-2}{2}-E \right)}{2^{d-1}\pi^{\frac{d+1}{2}}\Gamma\left(\frac{d-1}{2} \right)}\nonumber\\
&\times {}_2F_1\left(\frac{d-2}{2}+E,\frac{d-2}{2}-E,\frac{d-1}{2}, \frac{1-\cosh(s)}{2} \right)\ .
\end{align}

\subsubsection*{Regge limit of $D$-functions }
The result (\ref{bck1}) is sufficient to obtain the Regge limit of $\bar{D}$-functions. Using (\ref{dbar2}) we now obtain the leading Regge behavior 
\be
\bar{D}_{\Delta_1\Delta_2 \Delta_3 \Delta_4}(z,\bz)=i \(\frac{\pi^d}{2 \prod_i \Gamma(\Delta_i) }\)e ^{i\pi\frac{\Delta_{21}+\Delta_{34}}{2}} \sqrt{z\bz}\ f_{\Delta_1\Delta_2 \Delta_3 \Delta_4}\(\frac{1}{2}\log\(z/\bz\) \)\ .
\ee
The $\bar{D}$-functions completely determine the Regge limit of $D$-functions for all kinematics. In particular, for the kinematics (\ref{points}) in the limit (\ref{regge2}), we obtain
\be
D_{\Delta_1\Delta_2 \Delta_3 \Delta_4}(\r,\br)=i \frac{\pi^d2^{1-\sum_i \Delta_i}}{(\rho \bar{\rho})^{\frac{\Delta_1+\Delta_2}{2}}\prod_i \Gamma(\Delta_i) }\sqrt{\frac{\br}{\r}}\ f_{\Delta_1\Delta_2 \Delta_3 \Delta_4}\(-\frac{1}{2}\log\(\r\br\) \)\ .
\ee

\subsubsection*{Special case: $\Delta_1+\Delta_2=\Delta_3+\Delta_4$}
We now consider a special case $\Delta_1+\Delta_2=\Delta_3+\Delta_4=K+\frac{d}{2}$ which will be useful for us. In this case,
\begin{align}
f_{\Delta_1\Delta_2 \Delta_3 \Delta_4}(s)=\int_{-\infty}^\infty d\nu \Omega_{i\nu}(s)\Gamma\left(\frac{K+i\nu}{2} \right)^2 \Gamma\left(\frac{K-i\nu}{2} \right)^2\ . 
\end{align}
As explained earlier, we are only interested in the $\r\br\rightarrow 0$ limit of Regge conformal blocks which leads to further simplification. Using the integral representation of the hypergeometric function, one can show that in the limit $s\rightarrow \infty$
\be
f_{\Delta_1\Delta_2 \Delta_3 \Delta_4}(s)\approx 4  \pi^{1-\frac{d}{2}}   \Gamma(K) \Gamma\left(K+\frac{d}{2}-1 \right)e^{-\frac{1}{2}(-2+d+2K)s} s\ .
\ee
Therefore, the $D$-function in the Regge limit (\ref{regge2}), followed by $\r\br\rightarrow 0$ can be approximated as
\be\label{finalD}
D_{\Delta_1\Delta_2 \Delta_3 \Delta_4}(\r,\br)\approx -i \frac{\pi^{1+\frac{d}{2}}}{2^{2K+d-2} \r}\ \log\(\r\br\) \frac{\Gamma(K) \Gamma\left(K+\frac{d}{2}-1 \right)}{\prod_i \Gamma(\Delta_i) }\ .
\ee

\section{Triple-trace operators}\label{tripletrace}
Let us now construct triple-trace operators in CFT$_5$ from a scalar primary operator $\O$ with dimension $\Delta_\O=5$ in a generalized free theory.  It is easy to construct a spin-0 triple-trace operator from $\O$:
\be
[\O^3]_{n=0,\ell=0}=\frac{1}{\sqrt{6}} \O(x)\O(x)\O(x)\ 
\ee
which has dimension $\Delta^{(3)}(0,0)=15$. One can also easily check that there are no spin-1 triple-trace operator with $n=0$. However, it is easy to construct a unique spin-2 triple-trace operator $[\O^3]_{0,\ell=2}$: 
\be
[\O^3]_{n=0,\ell=2}=\lim_{x_2,x_3\rightarrow x_1}\frac{1}{\sqrt{768}} \left((\varepsilon.\partial_3)^2-\frac{6}{5} \varepsilon.\partial_2 \varepsilon.\partial_3 \right) \O(x_1)\O(x_2)\O(x_3)\ 
\ee
which has dimension $\Delta^{(3)}(0,2)=17$ and twist $\tau_2=15$. Note that in the above equation $\varepsilon$ is the null polarization vector associated with the spinning operator $[\O^3]_{n=0,\ell=2}$.

However, we are mostly interested with the spin-3 triple-trace operator $[\O^3]_{0,\ell=3}$. Consider the most general triple-trace operator (not necessarily a primary)
\begin{align}\label{app12}
\lim_{x_2,x_3\rightarrow x_1} \left(\sum_{i=1,2,3} a_i (\varepsilon.\partial_i)^3+ \sum_{i\neq j =1,2,3} b_{ij}(\varepsilon.\partial_i)^2 (\varepsilon.\partial_j)+c (\varepsilon.\partial_1)(\varepsilon.\partial_2)(\varepsilon.\partial_3) \right)\O(x_1)\O(x_2)\O(x_3)
\end{align}
and define
\be
A=\sum_{i=1,2,3} a_i\ , \qquad \sum_{i\neq j =1,2,3} b_{ij}=B\ .
\ee
When this spin-3 triple-trace operator is a primary, it must be orthogonal to both $[\O^3]_{n=0,\ell=0}$ and $[\O^3]_{n=0,\ell=2}$. This conditions fix $A$, $B$ and $c$ uniquely
\be
A=\sqrt{\frac{5}{651168}}\ , \qquad c=\frac{84}{25}A\ , \qquad B=-\frac{21}{5}A\ .
\ee
It is easy to check that the operator (\ref{app12}) with the above conditions, has the right two-point function of a spin-3 primary with dimension $\Delta^{(3)}(0,3)=18$ and twist $\tau_3=15$. Furthermore, one can show that two triple-trace operators that satisfy the above conditions cannot be orthogonal to each other implying there is a unique spin-3 triple-trace primary operator with twist $\tau_3=15$. 

Higher spin and higher twist triple-trace operators can be constructed in a systematic way by conglomerating operators following \cite{Fitzpatrick:2011dm}.

\end{spacing}


\end{document}